\documentclass[useAMS,usenatbib,referee]{biom}
\usepackage{amsmath}
\usepackage[figuresright]{rotating}
\usepackage{booktabs,amssymb,adjustbox}
\usepackage{pbox}
\usepackage[final]{pdfpages}
 \usepackage{afterpage}

\def\bSig\mathbf{\Sigma}

\title[A Generalized Levene's Scale Test]{A Generalized Levene's Scale Test for Variance Heterogeneity in the Presence of Sample Correlation and Group Uncertainty}

\author
{David Soave$^{1,2,*}$ and Lei Sun$^{3,1,**}$\\
$^{1}$Division of Biostatistics, Dalla Lana School of Public Health, University of Toronto,\\
Toronto, ON M5T 3M7, Canada\\
$^{2}$Program in Genetics and Genome Biology, Research Institute, The Hospital for Sick Children,\\ Toronto, ON M5G 0A4, Canada\\
$^{3}$Department of Statistical Sciences, University of Toronto, Toronto, ON M5S 3G3, Canada\\
$^{*}$\textit{email:} david.soave@mail.utoronto.ca\\
$^{**}$\textit{email:} sun@utstat.toronto.edu}

\begin{document}

%  This will produce the submission and review information that appears
%  right after the reference section.  Of course, it will be unknown when
%  you submit your paper, so you can either leave this out or put in 
%  sample dates (these will have no effect on the fate of your paper in the
%  review process!)

%\date{{\it Received January} 2016. {\it Revised February} 2016.  {\it Accepted March} 2016.}

%  These options will count the number of pages and provide volume
%  and date information in the upper left hand corner of the top of the 
%  first page as in published papers.  The \pagerange command will only
%  work if you place the command \label{firstpage} near the beginning
%  of the document and \label{lastpage} at the end of the document, as we
%  have done in this template.

%  Again, putting a volume number and date is for your own amusement and
%  has no bearing on what actually happens to your paper!  

%\volume{64}
%\pubyear{2016}
%\artmonth{March}

%  The \doi command is where the DOI for your paper would be placed should it
%  be published.  Again, if you make one up and stick it here, it means 
%  nothing!

%\doi{10.1111/j.1541-0420.2005.00454.x}

%  This label and the label ``lastpage'' are used by the \pagerange
%  command above to give the page range for the article.  You may have 
%  to process the document twice to get this to match up with what you 
%  expect.  When using the referee option, this will not count the pages
%  with tables and figures.  

\label{firstpage}

%  put the summary for your paper here

\begin{abstract}
We generalize Levene's test for variance (scale) heterogeneity between $k$ groups for more complex data, which includes sample correlation and group membership uncertainty.  Following a two-stage regression framework, we show that least absolute deviation regression must be used in the stage 1 analysis to ensure a correct asymptotic $\chi^2_{k-1}/(k-1)$ distribution of the generalized scale ($gS$) test statistic.  We then show that the proposed $gS$ test is independent of the generalized location test, under the joint null hypothesis of no mean and no variance heterogeneity. Consequently, we generalize the recently proposed joint location-scale  ($gJLS$) test valuable in settings where there is an interaction effect, but one interacting variable is not available.  We evaluate the proposed method via an extensive simulation study, and two genetic association application studies.  
\end{abstract}

%  Please place your key words in alphabetical order, separated
%  by semicolons, with the first letter of the first word capitalized,
%  and a period at the end of the list.
%

\begin{keywords}
Heteroscedasticity; Scale test; Joint location-scale test; Association studies.
%Variance Test; Correlated samples; Probabilistic data; Association studies.
\end{keywords}

%  As usual, the \maketitle command creates the title and author/affiliations
%  display 

\maketitle

%  If you are using the referee option, a new page, numbered page 1, will
%  start after the summary and keywords.  The page numbers thus count the
%  number of pages of your manuscript in the preferred submission style.
%  Remember, ``Normally, regular papers exceeding 25 pages and Reader Reaction 
%  papers exceeding 12 pages in (the preferred style) will be returned to 
%  the authors without review. The page limit includes acknowledgements, 
%  references, and appendices, but not tables and figures. The page count does 
%  not include the title page and abstract. A maximum of six (6) tables or 
%  figures combined is often required.''

%  You may now place the substance of your manuscript here.  Please use
%  the \section, \subsection, etc commands as described in the user guide.
%  Please use \label and \ref commands to cross-reference sections, equations,
%  tables, figures, etc.
%
%  Please DO NOT attempt to reformat the style of equation numbering!
%  For that matter, please do not attempt to redefine anything!

\section{Introduction}
\label{s:introduction}

Testing for scale (variance) heterogeneity, prior to the main inference of location (mean) parameters, is a common diagnostic method in linear regression to evaluate the assumption of homoscedasticity.  In some research areas, such as statistical genetics, testing for heteroscedasticity itself can be of primary interest.

With the goal of detecting a genetic association between a single-nucleotide polymorphism (SNP, $G$) and a quantitative outcome (phenotype, $Y$), the traditional approach is to conduct a location test, testing mean differences in $Y$ across the three genotype groups of the SNP ($G=0$, 1 or 2 copies of the minor allele, the variant with population frequency $<0.5$).  However, it has been noted that a number of biologically meaningful scenarios can lead to variance differences in $Y$ across the genotype groups of a SNP of interest (say $G_1$).  For example, an underlying interaction effect between $G_1$ and another SNP $G_2$ ($G_1$x$G_2$) or an environmental factor $E$ ($G_1$x$E$), on $Y$ can lead to heteroscedasticity across $G_1$ if the interacting $G_2$ or $E$ variable is not collected to directly model the interaction term \citep{RN25}.   Transformations on a phenotype can also result in variance heterogeneity \citep{RN95}.  This transformation can occur knowingly for statistical purposes, e.g.\ log($Y$), or unknowingly, e.g.\  choosing a phenotype measurement that does not directly represent the true underlying biological outcome of a gene.  In each of these scenarios, a scale test can be used either alone to indirectly detect associated SNPs \citep{RN25}, or combined with a location test to increase testing power \citep{RN118, RN212}.  

Genotype uncertainty is inherent in both sequenced and imputed SNP data.  For these types of data, the genotype of a SNP for an individual ($G=0$, 1 or 2) is represented by three genotype probabilities ($p_0$, $p_1$, $p_2$, and $p_0+p_1+p_2=1$).  For testing methods that require genotype to be known unambiguously, the probabilistic data are typically transformed into the so-called ``best-guess" (most likely or hard-call) genotype, crudely selected as the one with the largest probability.   In the context of location-testing, several groups have proposed methods that incorporate the probabilistic data and showed that this improves power \citep{RN140, RN196}.  The corresponding development for scale-testing, however, is lacking.

Genetic association studies often involve family data, where individuals in a sample are correlated or clustered.  In addition, unintentional correlation due to cryptic relatedness may be revealed from standard quality control analyses of a population sample of presumed unrelated individuals \citep{RN287}.  A number of generalized location tests allowing for family data have been proposed \citep{RN135, RN275}, and their power gain over analyzing only the subset of independent individuals is a direct consequence of the increase in sample size.  However, few scale tests deal with correlated data, with the exception of methods proposed specifically for clustered data present in twin studies \citep{RN197, RN182}.  Further, these methods have been reported to have  type 1 error issues in the presence of non-normal data or small, unequal group sizes \citep{RN182}, and they have not been extended to incorporate group membership uncertainty.
  
Both classical statistical tests and graphical procedures have been proposed to investigate heteroscedasticity \citep{RN193, RN278, RN176, RN27, RN194}.  In big data settings, such as genome-wide association studies, where possibly millions of SNPs are scanned for association with an outcome, graphical and other computationally burdensome approaches are undesirable.  Levene's test \citep{RN27} is known for its simplicity and robustness to modelling assumptions, and it is perhaps the most popular method for evaluating variance heterogeneity between $k$ groups.  Therefore, our development here focuses on Levene's method. 

In this paper, we extend Levene's test for equality of variances across $k$ groups to allow for both group membership uncertainty and sample correlation.  When groups are known, we show that the proposed method outperforms existing methods for clustered twin data.  In the presence of group uncertainty, we demonstrate that our test continues to be accurate and has improved power over the ``best-guess" approach.  This generalized scale test can be used alone for heteroscedasticity diagnostic purposes but with wider applicability.  Motivated by the complex genetic association studies described above, we also show that the proposed $gS$ test can be combined with existing generalized location tests using the joint location-scale framework, previously developed for population samples without group uncertainty \citep{RN212}, to further improve power.  Finally, we apply our methods to two genetic association studies, one of HbA1c levels in individuals with type 1 diabetes, and the other of lung disease in individuals with cystic fibrosis.
%Motivated by the complex genetic data as described above, we apply our methods to two genetic association studies, one on HbA1c levels in individuals with type 1 diabetes, and the other on lung disease in individuals with cystic fibrosis.

\section{Methodology}
\label{s:Methodology}

We first consider a sample of independent observations with no group uncertainty, and formulate Levene's test as a regression problem. Using this regression framework, we then extend Levene's test as the generalized scale ($gS$ hereinafter) test  to allow for sample dependency and group uncertainty.  For clarity of the methods comparison, we also briefly discuss the \cite{RN182} extension of Levene's test, specifically designed for twin pairs without group uncertainty.  Finally, we generalize the joint location-scale test of \cite{RN212} ($gJLS$) for the complex data structure considered here. 

\subsection{Notation and Statistical Model}

Let $y_i, i=1,\ldots,n$, be a sample of independent observations, where each $y_i\sim \mathcal{N}(\mu_i,\sigma_i^2)$.  Suppose the $y_i$'s fall into $k$ distinct treatment groups with group-specific variance $\sigma_j^2$, $j=1,\ldots,k$, and let $n_j$ be the sample size for group $j$, $n=\sum n_j$.  Our motivation concerns testing the hypothesis of equal variance across the $k$ groups:
\begin{equation}
H_0: \sigma_1^2=\sigma_2^2=\dots=\sigma_k^2.
\label{null1}
\end{equation}
For notation concision, here we use  $\sigma_j^2$  for group-specific variance, $j=1,\ldots,k$, and $\sigma_i^2$ for observation-specific variance, $i=1,\ldots,n$; in what follows we make the distinction clear in the context.

Let $x_{ji}, j=1,\ldots,k-1$, be the standard dummy variables, where $(x_{1i}=0,\ldots, x_{(k-1)i}=0)$ for observation $i$ belonging to group 1, and $(x_{1i}=0,\ldots, x_{(j-1)i}=1,\ldots,x_{(k-1)i}=0)$ for group $j$, $j=2,\ldots,k$.   
%In genetic association setting where $k=3$,  $(x_{1i}=0, x_{2i}=0)$ for individual $i$ with genotype 0, $(x_{1i}=1, x_{2i}=0)$ for genotype 1, and $(x_{1i}=0, x_{2i}=1)$ for genotype 2.  

Consider the normal linear model of interest here,
\begin{equation}
\begin{split}
y_i= \beta_0+\beta_1 x_{1i}+\beta_2 x_{2i}+\dots+\beta_{k-1} x_{(k-1)i}+\varepsilon_i,\\
i=1,\dots,n,
\end{split}
\label{yi.eq}
\end{equation}
where  $\varepsilon_i\sim \mathcal{N}(0,\sigma_i^2)$,  $\sigma_i^2$ corresponds to the variance associated with the group that $y_i$ belongs to.  In other words, $\sigma_i^2=\sigma_{j^*}^2$ if $x_{(j^*-1)i}=1$. 
In matrix notation,
\begin{equation}
\bm{y}= X\bm{\beta}+\bm{\varepsilon},
\label{yMatrix}
\end{equation}
where $X$ is the design matrix obtained by stacking the $\bm{x}_i^T=(1,x_{1i},x_{2i},\dots,x_{(k-1)i})$, $\bm{\varepsilon}\sim \mathcal{N}_n (\bm{0},\Sigma)$, and $\Sigma$ is the covariance matrix with diagonal elements $\sigma_i^2$s. 

\subsection{Formulating Levene's Test as a Regression F-test and Modifications}

%Levene's test [Levene 1960] is well known as a test of equality of variances across groups, testing the null hypothesis of (1).   
The classical formulation of Levene's test first centres the observations, $y_i$'s, by their estimated group means and obtains the corresponding absolute deviations, $d_i$'s.  It then tests for mean differences in the $d_i$'s across the $k$ groups using ANOVA.  Let $I_{ij}, j=1,\ldots,k$ be the group indicator variables, where $I_{ij}=1$ if individual $i$ belongs to group $j$.  Now, let $\overline{\mu_{(j)}}= \sum_{i=1}^n y_i I_{ij}/n_{j}$ be the estimated group means of the $y_i$'s, such that an estimate of $E(y_i)$ is $\overline{\mu_{i}}= \sum_{j=1}^kI_{ij}\overline{\mu_{(j)}}$. The corresponding absolute deviations are
$$d_i = |y_i - \overline{\mu_{i}}|.$$
Let $\overline{d_{(j)}}$ be the estimated group means of the $d_i$'s, such that an estimate of $E(d_i)$ is $\overline{d_i}= \sum_{j=1}^kI_{ij}\overline{d_{(j)}}$, and let $\overline{\overline{d}}={\sum_{i=1}^n d_i }/{n}$ be the grand mean.  Finally, Levene's test statistic has the following form
$$F(\bm d)=\frac{ \sum_{i=1}^n (\overline{d_{i}}-\overline{\overline{d}})^2/(k-1)} { \sum_{i=1}^n (d_{i}-\overline{d_i})^2/(n-k) },$$
where $F(\bm d)$ follows approximately an $F(k-1, n-k)$ distribution under the null hypothesis of (\ref{null1}), and a $\chi_{k-1}^2/(k-1)$ distribution asymptotically as $n \to \infty$.
 
For the purpose of a unified development, it is prudent to re-formulate Levene's test using the following two-stage regression framework: 
\begin{enumerate}[1]
\item[Stage 1.1.] Obtain the residuals, $\widehat{\varepsilon}_i=y_i-\widehat{y}_i=y_i - \bm{x}_i \widehat{\bm{\beta}} $, from the ordinary least squares (OLS) regression of $y_i$ on $\bm{x}_i^T$; we refer to this as the \textit{stage 1} regression.  
\item[Stage 1.2.] Take the absolute values of these residuals, $d_i$=$|\widehat{\varepsilon}_i|$.
\item[Stage 2.] Test for an association between the $d_i$'s and $\bm{x}_i^T$'s using a regression $F$-test; we refer to this as the \textit{stage 2} regression and test.
\end{enumerate}

The justification for this two-stage regression procedure (Levene's test) being a test of the hypothesis of variance homogeneity (\ref{null1}) is as follows.  Stage 1 performs OLS regression using a working covariance matrix $\Sigma_{stage\:1} = \sigma^2_{y} I$,  where $I$ is the identity matrix.  
Therefore $\bm{\widehat{y}}= X(X^T X)^{-1} X^T y=Hy$, 
$\bm{\widehat{\varepsilon}} = \bm{y} - \bm{\widehat{y}}  \sim \mathcal{N}(\bm{0}, \Sigma (I-H))$ and 
$\widehat{\varepsilon}_i \sim \mathcal{N}(0, \sigma_i^2 (1-h_{ii}))$,  where $h_{ii}$ is the $i$th diagonal element of the hat matrix $H$.
Consequently $d_i=|\widehat{\varepsilon}_i|$ follows a folded-normal distribution and its mean is a linear function of $\sigma_i$,
\begin{equation*} 
E(d_i )=\sigma_i \sqrt{\frac{2}{\pi} (1-h_{ii})}.
\end{equation*}
%This equation is exact for normal data and an approximation otherwise.  
This relationship between $d_i$ and $\sigma_i$ is approximated by the following working model in stage 2,
\begin{equation}
d_i= \alpha+\gamma_1 x_{1i}+\gamma_2 x_{2i}+\dots+\gamma_{k-1} x_{(k-1)i}+e_i,
\label{di.eq}
\end{equation}
where 
%$e_i \sim \mathcal{N}(\bm{0}, \Sigma_{stage\: 2}), \Sigma_{stage\:2}=\sigma^2_{d} \Sigma_d$. 
$e_i \sim \mathcal{N}(\bm{0}, \sigma^2_{d})$. 
In matrix form,
\begin{equation}
\bm{d}= X\bm{\theta}+\bm{e},
\label{dMatrix}
\end{equation}
where $\bm{\theta} = (\alpha, \bm{\gamma}^T)^T=(\alpha, \gamma_1, \dots, \gamma_{k-1})^T$, and 
$\bm{e} \sim \mathcal{N}(\bm{0}, \Sigma_{stage\: 2}), \Sigma_{stage\:2}=\sigma^2_{d} I$.
Testing the null hypothesis (\ref{null1}) is now re-formulated as testing
\begin{equation}
H_0: \gamma_1= \gamma_2=\dots= \gamma_{k-1}=0,
\label{null2}
\end{equation}
using the classical OLS regression $F$-test.  Note that although the $d_i$'s are folded normal variables, Levene's variance test takes advantage of the fact that inference from OLS regression is robust to violations of the normality assumption.  
%A discussion of the rescaling of $d_i$ by $(1-h_{ii})^{-1/2}$ is given in Section 5.

This formulation of Levene's test has a similar structure to the score test of \cite{RN186} proposed for testing heterscedasticity associated with continuous covariates.  \cite{RN203} showed that when estimating ${\bm{\beta}}$  by OLS in the stage 1 regression, the resulting Glejser score statistic derived from the stage 2 regression analysis is not asymptotically distributed as $\chi^2_1$, unless the distribution of $\bm{\varepsilon}$ is symmetric.  To achieve robustness, several modifications have been proposed \citep{RN185, RN204, RN205, RN201, RN214}, among which replacing sample group means with medians in constructing the $d_i$'s is most intuitive.  This substitution has been consistently recommended in the literature for its robustness against non-normality \citep{RN198, RN200}.  It has also been shown analytically that, when the distribution of the error $\bm{\varepsilon}$ is not symmetric, centering on the sample group medians, and not the means, will lead to an asymptotically correct Levene's test \citep{RN199} and correct Glejser's score test \citep{RN201}. In the regression framework, this modification corresponds to estimating ${\bm{\beta}}$ by least absolute deviation (LAD) regression instead of OLS regression in stage 1.

\subsection{The Generalized Levene's Scale ($gS$) Test}

The above regression framework for Levene's test allows us to incorporate group uncertainty by simply replacing the group indicators or dummy variables for each observation, $\bm{x}_i^T$, with the corresponding group probabilities.  Analogous to dummy variables, the group probabilities for each individual sum to 1, so we omit one of the covariates to ensure model identifiability.  Using genetic association as an example again, let 
$(p_0=0.25, p_1=0.42, p_2=0.33)$ be the genotype probabilities for an individual $i$ at a SNP of interest, then, without loss of generality, we can define  
$\bm{x}_i^T=(1,x_{1i},x_{2i})=(1, 0.42, 0.33)$. Note that the ``best-guess" approach would have the corresponding covariate vector as  $\bm{x}_i^T=(1, 1, 0)$.   

Now, consider correlated data where $\varepsilon_i$ and $\varepsilon_j$ are no longer independent of each other and the covariance matrix $\Sigma$ is no longer diagonal.  In the stage 1 regression, because we are only interested in obtaining $\widehat{\bm{\beta}}$ to construct  $d_i=|y_i - \bm{x}_i \widehat{\bm{\beta}}|$, we can continue to use OLS or LAD regression with the misspecified working covariance matrix,  $\Sigma_{stage\:1} = \sigma^2_{y} I$, to obtain consistent and unbiased  ${\bm{\beta}}$ estimates.

Stage 2 involves estimating the variance of $\widehat{\bm{\gamma}}$ to test the null hypothesis of (\ref{null2}), and not accounting for sample dependency  can lead to invalid inference. Let $\Sigma_{stage\: 2}=\sigma^2_{d} \Sigma_d$;
a valid inference can be achieved by using a generalized least squares (GLS) approach when $\Sigma_{d}$ is known \citep{RN279}.  When $\Sigma_{d}$ is unknown, feasible GLS (FGLS) \citep{RN277} can be used, with or without iteration, where an estimate of $\Sigma_{d}$ is obtained, subject to constraints, and then used in GLS.  Alternatively, orthogonal-triangular decomposition methods can be used to obtain a compact representation of the profiled log-likelihood, such that maximum likelihood estimates (MLE's) of all parameters can be obtained jointly through nonlinear optimization \citep{RN175}.  

In many scientific settings, including genetic association studies, the sample correlation structure is often specified with constraints on the $n(n-1)/2$ correlations, e.g. a single serial correlation $\rho$ for time series or family data with a single relationship type (e.g. twin data), or different cluster-specific correlations $\rho$'s for different clusters.  
In this case, 
%$\Sigma_{d}$ can be estimated by ML.  Briefly, 
let $\Sigma_{stage\: 2}=\sigma^2_{d} \Sigma_d(\rho)=\sigma^2_{d} C(\rho)C(\rho)^T$  be the Cholesky decomposition, and define 
$$\bm{d}^*=C(\rho)^{-1}  \bm{d}, \: \: X^*=C(\rho)^{-1}  X, \:  \bm{e}^*=C(\rho)^{-1}\bm{e}$$
The GLS or FGLS regression, in essence, deals with the transformed model in stage 2
\begin{equation}
\bm{d^*}= X^*\bm{\theta}+\bm{e^*},
\label{d*Matrix}
\end{equation}
where $\bm{\theta} = (\alpha, \bm{\gamma}^T)^T$.
For a fixed $\rho$, the conditional MLEÕs for $\bm{\theta}$ and $\sigma_d^2$ are
$$\widehat{\bm{\theta}}=[X^{*T} X^*]^{-1}X^{*T} \bm{d}^*,\: \: \widehat{{\sigma}_d^2}=\frac{1}{n} \left\Vert \bm{d}^*-X^* \widehat{\bm{\theta}} \right\Vert ^2.$$
The MLE of $\rho$ can be obtained by optimizing the profiled log-likelihood,
\begin{equation*}
l(\rho)=\text{constant}-n \text{log}\left\Vert \bm{d}^*(\rho)-X^*(\rho) \widehat{\bm{\theta}}(\rho)\right\Vert -\frac{1}{2} log|C(\rho)|.
\end{equation*}
%and then substitute $\widehat{\rho}$ to get the MLEÕs for $\bm{\theta}$ and $\sigma_d^2$.  Additional details of ML estimation, including numerically efficient representations of the profiled log-likelihood, are described in Pinheiro, et al. [2000].

Thus, the generalized Levene's scale $gS$ test of the null hypothesis of (\ref{null2}), $H_0: \bm{\gamma}=\bm{0}$, using the regression F-test in stage 2, has the following test statistic:
\begin{equation}
F(\bm{d}^*)=\frac{{\sum_{i=1}^{n}(\widehat{d_{i}^*}-\widetilde{d_{i}^*})^2}/{(k-1)}}{{\sum_{i=1}^{n}(d_{i}^*-\widehat{d_{i}^*})^2}/{(n-k)}},
%\begin{split}
%F(\bm{d}^*)=\frac{MSR(\bm{d}^* )}{MSE(\bm{d}^*)},\\
%MSR(\bm{d}^*)={\sum_i^{n}(\widehat{d_{i}^*}-\widetilde{d_{i}^*})^2}/{(k-1)},\\
%MSE(\bm{d}^*)={\sum_i^{n}(d_{i}^*-\widehat{d_{i}^*})^2}/{(n-k)},
%\end{split}
\label{Fstat.d}
\end{equation}
where $\widehat{d_{i}^*}=(\bm{x}_{i}^* )^T \bm{\widehat{\theta}}$, the predicted values from regression model (\ref{d*Matrix}), and $\widetilde{d_{i}^*}={1}_{i}^* {\widetilde{\alpha}}$, the predicted values from the regression of $\bm{d}^*$ on $\bm{1}^*$.
Note that $\bm{1}^*$ is the first column of the transformed design matrix $X^*$, and may not be a vector of $1$'s.  
When the observations are independent of each other and group membership is known unambiguously, it is easy to verify that $\widehat{d_{i}^*}=\overline{d_i}$ and $\widetilde{d_{i}^*}=\overline{\overline{d}}$, and  $F(\bm{d^*})$ reduces to the original form of $F(\bm{d})$. 

%Under the null hypothesis of homoscedasticity of (\ref{null2}), 
Under the linear regression model of (\ref{di.eq}), 
the $F$-statistic (\ref{Fstat.d}) of testing (\ref{null2}) is asymptotically $\chi_{k-1}^2/(k-1)$ distributed \citep{RN206}.   However, similar to the results of \cite{RN199} and \cite{RN201} for the original Levene's test, we show that for non-symmetric $\bm{\varepsilon}$, this is true only when $\bm{d}$ is estimated using LAD in the stage 1 regression  (Web Appendix A, Theorem 1).

\subsection[subsection heading]{The \cite{RN182} Scale Test for Twin Pairs and Modifications}

Focusing on paired-observations, \cite{RN182} extended Levene's test to determine if the variance of an outcome differs between monozygotic (MZ) and dizygotic (DZ) twin pairs.  The proposed twin ($TW$) test follows Levene's two-stage regression procedure but it makes use of the Huber-White sandwich estimate \citep{RN194} of Var$(\widehat \gamma_1)$ in the stage 2 analysis (here $k=2$ requiring only one dummy variable) to construct an asymptotically $\chi_1^2$ distributed Wald statistic, operationally an $F$-statistic in finite samples.   

Complications with the $TW$ test may arise if the number of clusters is small in either group (MZ or DZ) %\citep{RN209}, 
and can be compounded with imbalance between the groups \citep{RN182}.  Unfortunately, there is no clear definition of too few clusters \citep{RN209}, and empirical type 1 error rates can be inflated for study designs with less than 20 clusters per group, particularly combined with non-symmetric data (see \cite{RN182} and simulation results Section 3 below).  
%More importantly, this approach assumes that if two observations are from the same pair/cluster, then they belong to the same group $k$ with $\sigma_{k}^2$.  This restriction may not be satisfied in a more general setting like the genetic association studies discussed above.  For example, two individuals from the same DZ pair or familial cluster often have different genotypes at a SNP of interest, so  a pair of individuals cannot be used to estimate a common $\sigma_k^2$.  {\bf ********* This statement is risky, there is no reason that a sandwich estimator to adjust for clustering couldn't be used for regression on a continuous covariate (i.e. the covariate would differ between the siblings in any cluster similar to discordant genotypes) }    

The original $TW$ method assumes that if two observations are from the same pair/cluster they also belong to the same group $k$.  This may not be satisfied in a more general setting like the genetic association studies discussed above.  For example, two individuals from the same DZ pair or familial cluster often have different genotypes at a SNP of interest, so individuals from the same cluster may not share a common $\sigma_k^2$.   However, the sandwich variance estimator can continue to be used in this setting.
In the presence of group uncertainty,  the $TW$ method can be modified by replacing the group indicator covariate with group probabilities. 
%allowing a comparison of $TW$ and $gL$ for the analysis of genotypes, incorporating group uncertainty.  The $TW$ test was originally implemented using the STATA statistical software command $regress$ using the $cluster()$ option to incorporate the cluster information in the standard error adjustment.  This method of cluster robust inference can also be implemented in the R software[R Core Development Team 2013. R Foundation for Statistical Computing, Vienna, Austria] (Web Appendix C).

\subsection{Generalized Joint Location-Scale (gJLS) testing}

The standard location test of mean differences in an (approximately) normally distributed outcome across covariate values (e.g.\ the three genotype groups of a SNP in a genetic association study) is testing
$$H_0^{location}: \beta_1=\ldots=\beta_{k-1}=0,$$ 
based on regression model (\ref{yi.eq}).  
%This is the stage 1 model used for the two-stage Levene's scale test of variance equality.
While the location test performs a hypothesis test on the $\beta_j$'s, the scale test discussed here uses only the $\beta$ estimates from the stage 1 regression of model (\ref{yi.eq}) to obtain $d_i=|y_i-\widehat y_i|$ for the stage 2 regression of model (\ref{di.eq}), and it performs a hypothesis test on the $\gamma_j$'s, testing
$$H_0^{scale}: \gamma_1=\ldots=\gamma_{k-1}=0.$$ 
A joint location-scale ($JLS$) test  is interested in the following global null hypothesis,
\begin{equation} 
%\begin{split} 
%H_0^{joint}: \beta_1=\ldots=\beta_{k-1}=0, \text{ and } \sigma^2_{1}=\ldots=\sigma^2_{k},\\
%or \\
H_0^{joint}: \beta_j=0, \text{ and } \gamma_{j}=0, \forall \hspace{2mm} j=1,\ldots,k-1.
%\end{split}
\label{nullJoint}
\end{equation}
%{\bf [Note to ourselfs: testing $\sigma$ and $\sigma^2$ may not be the same!]} 

One simple yet powerful $JLS$ method proposed in \cite{RN212} uses Fisher's method to combine $p_L$ and $p_S$, the $p$-values of the individual location and scale tests.  One can consider other aggregation statistics, e.g. the minimal $p$-value \citep{RN257, RN177}; for a review of this topic see \cite{RN97} and \cite{RN80}.  Focusing on Fisher's method,
the corresponding test statistic is
$$W_F=-2(log(p_L )+log(p_S)).$$   
For independent observations with no group uncertainty, \cite{RN212} showed that, under $H_0^{joint}$ of (\ref{nullJoint}) and a Gaussian model,
$p_L$ and $p_S$ are independent.  Thus $W_F$ is distributed as a $\chi_4^2$ random variable.

In the presence of sample correlation with group uncertainty, we propose to use the same framework but obtain $p_L$ from a generalized location test (e.g.\ a generalized least squares approach to model (\ref{yi.eq}), where the design matrix $X$ includes the group probabilities, and the covariance matrix, $\Sigma_{stage\:1} = \sigma^2_{y}\Sigma_{y}$, incorporates the sample correlation), and $p_S$ from the $gS$ test proposed here.  We show that the assumption of independence between $p_L$ and $p_S$ continues to hold theoretically under $H_0^{joint}$  of (\ref{nullJoint}) for normally distributed outcomes (Web Appendix B), as well as empirically for approximately normally distributed outcomes in finite samples (Web Figure 1).  

\section{Simulations}
\label{s:Simulations}
The validity of the generalized joint location-scale ($gJLS$) testing procedure  relies on the accuracy of the individual generalized location ($gL$) test and generalized scale ($gS$) test components.  The performance of the $gL$ test has been established in the literature, therefore, our simulation studies here focused on evaluation of the proposed $gS$ test, and when appropriate compared it with Levene's original test ($Lev$) and the $TW$ test of \cite{RN182}.  We use subscripts $_{OLS}$ and $_{LAD}$ to denote if the stage 1 regression was performed using OLS to obtain group-{\it mean}-adjusted residuals  or LAD for group-{\it median}-adjusted residuals.
Implementation details of each of the six tests ($Lev_{OLS}$, $Lev_{LAD}$, $TW_{OLS}$, $TW_{LAD}$, $gS_{OLS}$,$gS_{LAD}$)  is outlined in Web Appendix C. 

We considered two main simulation models.  Simulation model 1 followed the exact simulation setup of \cite{RN182} to ensure fair comparison.
Simulation model 2 extended model 1 by introducing genotype groups for each individual as well as group membership uncertainty.
% the $TW$ test is no longer applicable in this setting because paired individuals no longer belong to the same group.  
To apply the original $Lev$ test for comparison, we ignored the inherent sample correlation in the presence of correlated data.
In all simulations, empirical type 1 error and power were evaluated at the 5$\%$ significance level using 10,000 replicates, unless otherwise stated.

\subsection{Simulation Model 1}

\subsubsection{Model Setup}

Following the exact simulation study design of \cite{RN182}, we simulated correlated outcome values for $n_1$ MZ twin pairs and $n_2$ DZ twin pairs, $n=2n_{1}+2n_{2}$, and we tested if the variance of the outcome differed between the two groups of pairs, i.e. $\sigma_1^2=\sigma_2^2$.  To study robustness, we simulated outcomes using Gaussian, StudentÕs $t_4$ (heavier tailed), and $\chi_4^2$ (non-symmetric) distributions.  

We first generated pairs of observations from independent bivariate normal distributions $BV\mathcal{N}(0,1,\rho_k ), k=1,2$, with $\rho_1$ and $\rho_2$ corresponding to the correlation within the MZ and DZ twin pairs, respectively.  Let $w$ be the variable for an observation, we then applied a transformation $g(\cdot)$ to $w$ to obtain the desired marginal distribution, $y=\sigma_k g(w)$, where the $\sigma_k$'s induced different variances between the two groups.  The choice of $g(\cdot)$ depended on the desired distribution for $y$: 
\begin{gather*}
g(w)=
\begin{cases}
w, & \text{if } y \sim \mathcal{N}(0,1)\\
F_{t_4}^{-1} (\Phi(w)), & \text{if } y \sim t_4\\
F_{\chi_4^2}^{-1} (\Phi(w)), & \text{if } y\sim \chi_4^2
\end{cases}
,
\end{gather*}
where $\Phi$, $F_{t_4}$  and $F_{\chi_4^2}$ are the cumulative distribution functions for the standard normal, StudentÕs $t_4$ and $\chi_4^2$ distributions, respectively.

We varied the sample size ($n_1,n_2=5$, 10 or 20 for small samples, and $=500$, 1000 or 2000 for large samples, and $n_1$ may or may not equal $n_2$), and group variances ($\sigma_1^2, \sigma_2^2=1$, 2 or 4).  The level of correlation within the MZ and DZ twin pairs was $\rho_1=0.75$ and $\rho_2=0.5$, respectively.   %Further to these combinations considered by Iachine, et al. [2010], we also examined methods performance when not all individuals are paired.  Specifically, we considered additional data types where 50$\%$ of the sample clusters are singletons (and 50$\%$ are pairs), 90\% singletons, or all are singletons (the last case is equivalent to $\rho_1=\rho_2=0$ examined by Iachine, et al. [2010]).   

\subsubsection{Results}

We were able to replicate the simulation results of \cite{RN182} that studied 
$Lev_{OLS}$, $Lev_{LAD}$, $TW_{OLS}$, and $TW_{LAD}$ (Table 1 and Web Table 1).  However, we noticed that results reported in their paper for $Lev_{LAD}$ and $TW_{LAD}$ using median-adjusted residuals (labeled as $W_{50}$ and $TW_{50}$, columns 9 and 12 of Tables 1-4 in \cite{RN182}) were mistakenly replaced by the $Lev$ and $TW$ results obtained using 10$\%$ trimmed mean-adjusted residuals (labeled as $W_{10}$ and $TW_{10}$ in \cite{RN182}).  
%The results of $W_{50}$ and $TW_{50}$ reported in Iachine, et al. [2010] in fact matched closely our simulation results using  10\% trimmed means, while the results of $W_{10}$ and $TW_{10}$ matched closely our $TW_{LAD}.   
Subsequent conclusions in \cite{RN182} that the $TW$ method using the 10$\%$ trimmed mean ``performed best", therefore, are incorrect and should instead refer to $TW_{LAD}$ using median-adjusted residuals from the stage 1 regression.  

Our results in Table 1 clearly show that
\begin{itemize}
\item In the presence of sample correlation, Levene's original method $Lev$ that ignores the correlation had severely increased type 1 error rate, even with Gaussian data.  That is, $TW$ and $gS$ performed better than $Lev$.
\item When the error structure was non-symmetric ($\chi^2_4$) or the group sizes were small (e.g.\ $n_1$ or $n_2$ less than 20), using OLS in the stage 1 regression for either $TW$ or $gS$ led to increased type 1 error.  That is, $TW_{LAD}$ and $gS_{LAD}$ performed better than $TW_{OLS}$ and $gS_{OLS}$, respectively.
\item When the group sizes were unbalanced and small (e.g.\ $n_1=10, n_2=20$), $TW_{LAD}$ had increased type 1 error, even with Gaussian data.  That is, $gS_{LAD}$ performed better than $TW_{LAD}$.
\end{itemize} 

In large samples, the original $Lev$ test remained too optimistic, with an empirical $\alpha$ of 0.097 when $n_1=n_2=2000$ with Gaussian data (Web Table 1). The accuracy of both $TW_{LAD}$ and $gS_{LAD}$ increased as sample size increased, with empirical $\alpha$ of 0.052 when $n_1=n_2=2000$, even for the non-symmetric $\chi^2_4$ data.  The accuracy of both $TW_{OLS}$ and $gS_{OLS}$ also improved as sample size increased, however, only for symmetric Gaussian or $t_4$ data.  For $\chi^2_4$ data, their empirical  $\alpha$ level remained as high as $0.103$ when $n_1=n_2=2000$; this empirical result is consistent with Theorem 1 (Web Appendix A).

Because most of the six tests did not have good type 1 error control in the presence of sample correlation, small samples, unbalanced group sizes, or non-symmetric data, we delay the discussion of power until simulation model 2 below where we focus on methods comparison between $TW_{LAD}$ and $gS_{LAD}$, and in a more general simulation set-up. 

\subsection{Simulation Model 2}

\subsubsection{Model Setup}

The second simulation setup was motivated by genetic association studies as previously discussed.  We again considered sibling pairs to introduce sample correlation.  However, unlike simulation model 1, here we allowed individuals from the same pair/cluster to belong to different groups, where the groups were the different genotypes of a SNP of interest.  

Consider a SNP of interest with minor allele frequency (MAF) of $q$ ($=0.2$ or 0.1), we first simulated genotypes for  $n/2$ ($=20$, 50, 100, 500 or 1000) pairs of siblings.  To account for the inherent correlation of genotypes between a pair of siblings, we started with drawing the number of alleles shared identical by decent (IBD), $D=0$, 1 or 2,  from a multinomial distribution with parameters (0.25, 0.5, 0.25), independently for each sib-pair.  Given the IBD status $D$, we then simulated paired genotypes $(G_1, G_2)=(i, j), i, j \in \{0, 1, 2\}$, following the known conditional distribution of $\{(G_1, G_2)|D\}$ \citep{RN282, RN283}.  The distribution depends on $q$ in a way that smaller $q$ leads to greater imbalance in the genotype group sizes.  Approximately, the distribution of the numbers of individuals with genotype $G=0$, 1 and 2 is proportional to $(1-q)^2$, $2q(1-q)$ and $q^2$, respectively.

To introduce group membership uncertainty, we converted the simulated true genotypes $G$'s to probabilistic data $X$'s using a Dirichlet distribution. We used scale parameters $a$ for the correct genotype category and $(1-a)/2$ for the other two; this error model was used previously by \cite{RN140} to study location tests in the presence of genotype group uncertainty. We varied $a$ from 1 to 0.5, where $a=1$ corresponds to no genotype uncertainty and $a=0.5$ implies that, on average, 50\% of the ``best-guess" genotypes correspond to the true genotype groups.  Thus, the genotype group uncertainty level ranged from  0$\%$ to 50$\%$ in our simulations.

We then simulated outcome data for each sib-pair in a fashion similar to simulation model 1.  For each of the $n/2$ sib-pairs, 
we first simulated paired data from $BV\mathcal{N}(0,1,\rho)$, where $\rho=0.5$ was the within sib-pair correlation.  For each simulated value $w$, we then applied the $\sigma_k g(w)$ transformation to obtain the desired outcome data $y$ as in simulation model 1 (Gaussian, Student's $t_4$, and $\chi_4^2$).  However, $k$ here refers to the corresponding true underlying genotype group of an individual, and two individuals from the same sib-pair might not have the same genotype.  We used $(\sigma_0^2, \sigma_1^2, \sigma_2^2)=(1, 1, 1)$ to study type 1 error control, and $(1, 1.5, 2)$ or $(1, 2, 4)$ to study power; other values such as $(2, 1.5, 1)$ and $(4, 2, 1)$ were also considered.

It is evident from the results of simulation model 1 that the original $Lev$ test is not valid in the presence of sample correlation, and $TW_{OLS}$ and $gS_{OLS}$ are inferior, respectively, to $TW_{LAD}$ and $gS_{LAD}$,  when the error structure is non-symmetric or the group sizes are small. Therefore, the results presented below focus on comparison between $TW_{LAD}$ and  $gS_{LAD}$.  In the presence of genotype group uncertainty, we also considered the ``best-guess" approach and used $TW_{LAD}^{BG}$ and  $gS_{LAD}^{BG}$ to represent the corresponding results.

\subsubsection{Results}

In the presence of sample correlation but with no group uncertainty, the results in Table \ref{table2} show that both $TW_{LAD}$ and  $gS_{LAD}$ were accurate in large samples, e.g.\ when sample size was 2000 ($n/2=1000$ sib-pairs). However, $TW_{LAD}$ had increased type 1 error when group sizes were unbalanced and relatively small, even for Gaussian data. For example, when the MAF is $q=0.2$ and the number of sib-pairs is $n/2=100$, the expected  sizes of the three genotype group sizes are $n*((1-q)^2, 2q(1-q), q^2)=(128, 64, 8)$. In that case, the empirical type 1 error of $TW_{LAD}$ was 0.060, 0.072 and 0.078 for Gaussian, $t_4$ and $\chi^2_4$ data, respectively.  The problem was exacerbated by a smaller MAF $q=0.1$ with empirical type 1 error levels of 0.092, 0.115 and 0.118, respectively for the three types of data.
%by symmetric but non-Gaussian $t_4$ data (0.066 and 0.101, respectively for MAF of 0.2 and 0.1), and by non-symmetric $\chi^2_4$ data (0.072 and 0.116, respectively for MAF of 0.2 and 0.1).  
In contrast, the proposed $gS_{LAD}$ test remained accurate in most cases and was slightly conservative in small samples, when $n/2<100$. 
%n=200
%q=0.2
%print( c(n*(1-q)^2, n*2*q*(1-q), n*q^2))

Results in Table \ref{table3} are characteristically similar to those of Table \ref{table2}.  However, we note that group uncertainty somewhat mitigates the problem of unbalanced group sizes, and consequently the accuracy issue of $TW_{LAD}$. Nevertheless, it is clear that  $gS_{LAD}$ had better type 1 error control than $TW_{LAD}$ across the MAF values and the three outcome distributions.  As expected, $TW_{LAD}^{BG}$ and  $gS_{LAD}^{BG}$ using the ``best-guess" genotype group have similar type 1 error control to $TW_{LAD}$ and  $gS_{LAD}$ incorporating the group probabilistic data, under the null hypothesis (Table \ref{table3} and Web Tables 2 and 3).

Focusing on the accurate $gS_{LAD}$ test, Table 4 and Figure 1 demonstrate the gain in power when incorporating the group probabilistic data into the inference ($gS_{LAD}$) as compared to using the ``best-guess" group ($gS_{LAD}^{BG}$).  For example, at the 30\% group uncertainty level with sample size of 1000 ($n/2=500$ sib-pairs), MAF of 0.1 and under Gaussian data, the power of $gS_{LAD}$ was 0.613, a 23\% increase over the power of 0.495 observed for $gS_{LAD}^{BG}$; a similar gain in efficiency was observed for other sample sizes, MAF, and with $t_4$ and $\chi_2^4$ data (Table \ref{table4}).  

One would expect the relative efficiency gain to increase as uncertainty level increases. However, this is true only if the uncertainty level is not too high.  Depending on the model used to induce group uncertainty and the heteroscedasticity alternatives, it is reasonable to assume that the absolute power eventually converges to the type 1 error as the uncertainty increases. Consequently, the gain in relative efficiency of $gS_{LAD}$ compared to $gS_{LAD}^{BG}$ would also diminish and converge to 1.
%empirical estimate of the relative efficiency would also subject to higher variation.  
This is consistent with results in Figure 1.

\section{Applications}
\label{s:Applications}

To demonstrate the utility of the proposed generalized scale ($gS$) test and subsequent generalized joint location-scale  ($gJLS$) test, we revisited the two genetic association studies considered in \cite{RN212}, and compared our results with those using only a sample of unrelated individuals with no genotype group uncertainties.  We also used application data combined with simulation methods to further empirically validate the performance of the proposed methods.

\subsection{HbA1c Levels in Subjects with Type 1 Diabetes}

We use this application to demonstrate the gain in power by incorporating group uncertainty (probabilistic) data.  
Details of this dataset were previously reported in \cite{RN212}. Briefly, the outcome of interest was 
%averaged (over quarterly measured values spanning 6.5 years), 
inverse normal transformed HbA1c levels in $n=1304$ unrelated subjects with type 1 diabetes, and the SNP of interest was rs1358030 near \textit{SORCS1} on chromosome 10 with MAF of 0.36. 
%and the group sizes ($n_0, n_1, n_2$) are (545, 574, 183): two individuals with missing data at this SNP, then we have the problem of imputation. Better not to mention it.  
With no sample correlation or group uncertainty,  the original $Lev$ test for variance heterogeneity was applied and resulted in a significant result with $p=0.01$ \citep{RN212}.  Combined with other evidence reported in \cite{RN84}, we assume here that the association is real and smaller p-values implies greater power.

To demonstrate the effect of genotype group uncertainty, we masked the true genotypes using the same Dirichlet distribution as in the simulation studies above, where the value of $a$ ranged from 1 to 0.5, corresponding to no group uncertainty to 50\% uncertainty.  We then applied $gS_{LAD}^{BG}$ to the ``best-guess" genotype data and the proposed $gS_{LAD}$ incorporating the probabilistic data, and obtained the corresponding p-values, $p_{gS_{LAD}^{BG}}$  and $p_{gS_{LAD}}$.  For a given uncertainty level, we repeated the masking process independently 1,000 times and obtained averaged p-values on the log10 scale 
($10^{\{\text{average of } log10(p)\}}$),  $\overline{p}_{gS_{LAD}^{BG}}$  and $\overline{p}_{gS_{LAD}}$.
Between the two methods, it was clear that $gS_{LAD}$ was more efficient than $gS_{LAD}^{BG}$.  For example, when $a=0.75$ for 25\% group uncertainty, the $gS_{LAD}$ test remains significant with $\overline{p}_{gS_{LAD}}=0.048$ as compared to $\overline{p}_{gS_{LAD}^{BG}}=0.068$.  
%1.75, 1.67
%1.48, 1.36
%1.32,1.17 -> 0.048, 0.068
%1.19, 1.04  -> 0.065, 0.091
%x=1.48
%x=1.36
%10^(-x)
Regardless of the method used, the power of the scale tests decreased sharply as genotype uncertainty increased, consistent with those for location tests reported in \cite{RN140}, where location tests incorporating group uncertainty were compared with the ``best-guess" approach. 
%This difference increased (decreased) as the group uncertainty level increased (decreased), when the group uncertainty level is not too high as discussed previously. These results are also consistent with those for location tests reported in Acar and Sun [2013], where location tests incorporating group uncertainties were compared with the ``best-guess" approaches. 
% No need for (Web Figure 1)

\subsection{Lung Disease in Subjects with Cystic Fibrosis}

We used this application to demonstrate the gain in power by incorporating all available subjects, including relatives.
We also used this dataset combined with permutation methods to further demonstrate the validity of the proposed methods.
Details of this dataset were previously reported in \cite{RN212}. Briefly, the outcome of interest was lung function
%measured by SaKnorm (normalized, averaged sex-, age-, height-, mortality-adjusted forced expiratory volume in 1 s) 
as measured by the normally distributed SaKnorm quantity 
\citep{RN116} in a total of $n_{all}=1507$ individuals with CF, among which 1313 were singletons, 188 from 94 sib-pairs, and 6 from 2 sib-trios.  In total, 8 SNPs from 3 genes (\textit{SLC26A9}, \textit{SLC9A3} and \textit{SLC6A14}) were analyzed based on association evidence for other CF-related outcomes reported in \cite{RN49} and \cite{RN119}.    

Focusing on the $n_{indep}=1313+94+2=1409$ unrelated individuals, \cite{RN212} analyzed the association between lung function and each of the 8 SNPs using the individual location test and scale test, as well as the joint location-scale ($JLS$) test. They reported that SNPs from \textit{SLC9A3} and \textit{SLC6A14} were associated with CF lung functions (Table \ref{table5}).

The number of omitted subjects in that analysis was small ($n_{omit}=94+2*2=98$) and consequently the expected loss of efficiency is anticipated to be small.  Nevertheless, we re-analyzed the data available from the whole sample of $n_{all}=1507$ individuals, using the individual generalized location ($gL$) test, the proposed generalized scale ($gS$) test, and the subsequent generalized joint location-scale ($gJLS$) test (Table \ref{table5}).  We used a compound symmetric correlation structure (a single correlation parameter $\rho$) to model within family dependence for each application of GLS regression.

We first note that the conclusions for the presumed null SNPs from \textit{SLC26A9} did not change, as desired.  The conclusions for the presumed associated SNPs from  \textit{SLC9A3} and \textit{SLC6A14} did not change either, but using all available data led to smaller p-values for the $gL$ test.  The lack of apparent efficiency gain for $gS$
was somewhat disappointing, but it was also expected given the few number of siblings added to the sample; see Discussion Section 5 for additional comments. 
Lastly, we note that the $JLS$ framework indeed yields increased power when aggregating evidence from the individual tests; see \cite{RN212} for detailed discussions of the motivation and merits of the joint-testing framework.
  
To further exam the accuracy of the proposed $gS$ and $gJLS$ tests (as well as the $gL$ test for completeness), we generated 10,000 permutation replicates of the outcome to assess the empirical type 1 error control; permutation was performed separately between singletons and between sib-pairs; see \cite{RN281} for permutation techniques for more general family data.  Without loss of generality, we focused on SNP rs17563161 from \textit{SLC9A3} (Web Figure 1).   Testing the resulting $p$-values for deviation from the expected Uniform(0,1) distribution using the Kolmogorov-Smirnov test showed that all tests were valid.  
%with the KS p-values of XXX, XXX, 0.47, respectively for the $gL$, $gS$, and $gJLS$ tests. 
Additional simulations for inducing genotype group uncertainty led to the same conclusion (results not shown).

\section{Discussion}
\label{s:Discussion}

Levene's scale test is widely used as a model diagnostic tool in linear regression, and more recently it has been employed as an indirect test for interaction effects. Increased data complexity due to sample correlation or group uncertainty, however, limits its applicability.  Here we proposed a generalization of Levene's scale test, $gS$, that has good type 1 error control in the presence of sample correlation, small samples, unbalanced group sizes, and non-symmetric outcome data.  We showed that the least absolute deviation (LAD) regression approach to obtain group-{\it median}-adjusted residuals is needed to ensure robust performance of $gS$. Based on our results, we recommend the use of $gS_{LAD}$ over $gS_{OLS}$ (and other existing tests) uniformly for all study analyses.

\begin{figure}
  \centerline{\includegraphics[width=6in,trim={1 1.8cm 0cm 0cm},clip]{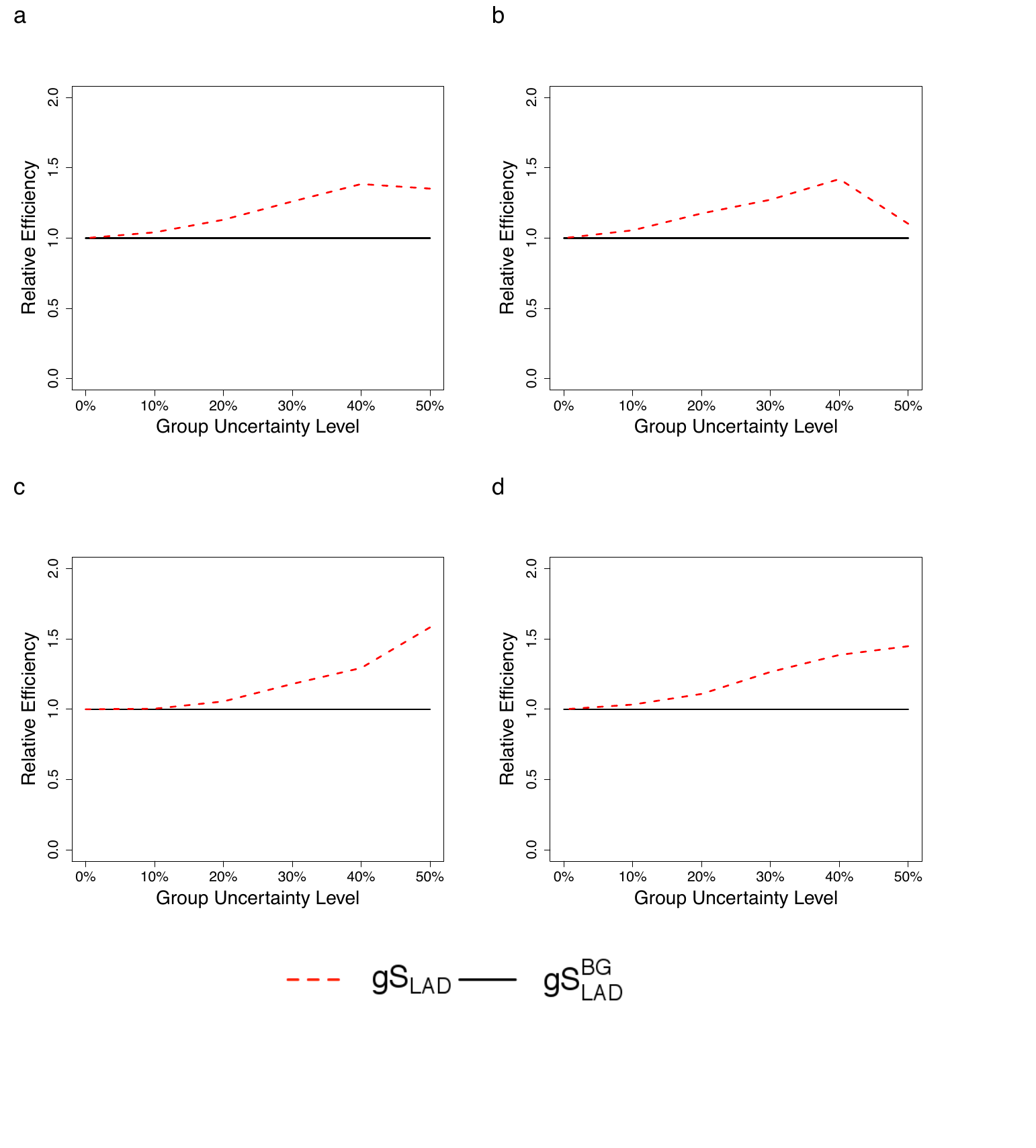}}
  \vspace*{1 cm}

\caption{\small{\textbf{Relative efficiency of $gS_{LAD}^{BG}$ and $gS_{LAD}$ under simulation model 2.} $gS_{LAD}^{BG}$ denotes $gS_{LAD}$ applied to the ``best-guess" genotype data, and the relative efficiency is the power of $gS_{LAD}$ divided by the power of $gS_{LAD}^{BG}$.  The true genotypes were masked using a Dirichlet distribution with scale parameters $a$ for the correct genotype and $(1-a)/2$ for the other two. On average, $a=0.5$ corresponds to 50\% group uncertainty, and $a=1$ corresponds to no genotype uncertainty (i.e. 0\%).  Parameter values included  $n/2=500$ sib pairs, and $\rho=0.5$ within-pair correlation.  The MAF and 
$(\sigma_0^2,\sigma_1^2,\sigma_2^2)$ were 0.1 and (1, 1.5, 2) for (\textbf{a}),  0.1 and (2, 1.5, 1)  for (\textbf{b}), 0.2 and (1, 1.5, 2) for (\textbf{c}), and 0.2, (2, 1.5, 1) for (\textbf{d}).  Power was estimated from 1,000 simulated replicates at the 5\% level. 
The  absolute power of $gS_{LAD}^{BG}$ and $gS_{LAD}$ at the 30\% group uncertainty level for (\textbf{a}) and (\textbf{c}) are presented in Table \ref{table4} under Gaussian data.}}

\label{f:figure1}
\end{figure}

In the presence of group membership uncertainty, $gS$ incorporating the probabilistic data increases power compared to using the ``best-guess" group data.  However, based on the simulations considered here, we note that when the group uncertainty level is moderate (e.g. 30\%), the efficiency gain is also moderate (Table \ref{table4} and Figure \ref{f:figure1}). When the group uncertainty is too high, the relative efficiency gain may diminish because the absolute power decreases considerably and eventually converges to the the type 1 error rate.  
%This conclusion does depend on mechanism of uncertainty (Dirichlt), but for genetic association studies considered here, we expect the ``best-guess" approach to work satisfactorily in practice.  

In the presence of sample correlation, the original $Lev$ test is inadequate due to inflated type 1 error.  Using a subset of only unrelated individuals would improve the accuracy of $Lev$ but at a cost to the power. The size of the efficiency loss depends on the proportion omitted from the sample as well as the dependency structure.  The $TW$ method of \cite{RN182} extends the $Lev$ test for twin data. Their simulation study as well as ours showed that $TW$ has an increased type 1 error rate when group sizes are unbalanced and relatively small, in contrast to the proposed $gS$.  When all group sizes were large, $gS$ and $TW$ were empirically equivalent. 

In the CF application, although the $gS$ test yielded comparable or less significant results after reincorporating siblings in the analysis, we observed that the corresponding $gL$ test results were more significant.  We considered the possibility that even though scale differences existed in the data, the addition of only 98 siblings (7$\%$ increase from the independent sample) may not yield a noticeable improvement in power of the $gS$ test.  Using the setup of simulation model 1, we examined the effect of incorporating only a small proportion of additional related subjects to an otherwise independent sample (Web Table 4).  We found that, compared with using a sample of $1000$ singletons, using a sample of $n=900$ singletons along with $100$ sib-pairs (10\% increase) led to a $<5\%$ power increase. In contrast, the addition of siblings to all unrelated subjects provided a substantial increase in power (Web Table 4). These results, and the noticeable power gain from the $gL$ location test when applied to the same CF data, are consistent with previous observations in genetic association studies that, larger samples are needed to detect variance differences as compared to mean differences \citep{RN216, RN215}.
%We found that addition of siblings to 10$\%$ of a sample of 1000 unrelated individuals lead to a negligible increase in power, whereas addition of siblings to all unrelated subjects provided a substantial increase in power (Web Table 10).  These results, and the noticeable gain from the $gL$ location test applied to the CF data, are consistent with previous observations in genetic association studies that larger samples are needed to detect variance differences as compared to mean differences [Visscher and Posthuma 2010; Yang, et al. 2011].
% This was supported by our additional simulation studies comparing power between using a sample of $1000$ singletons and using a sample of $n=900$ singletons along with $100$ sib-pairs; power difference was less than 5\%. The same amount of sample size increase, not surprisingly, can have bigger impact on testing location than scale parameters as seen here. In settings where most samples are related (e.g.\ $n/2$ sib-pairs), we expect substantial increase in power of $gS$ by using all available $n$ individuals as compared to power of $Lev$ using only the $n/2$ unrelated individuals; results thus are not presented here. 
  
The examination of the proposed $gS$ here focused on SNP genotype categories.  The so-called ``additive" coding of the genotype data can be used in practice. That is, replacing the two dummy variables, $x_1$ and $x_2$, with one continuous variable coded as $x=0, 1$ or 2, if there is no group uncertainty; or replacing the two probabilistic variables, $x_1$ and $x_2$, with an expected count (the so-called ``dosage"), $x=p_1+2*p_2$.  If the underlying model is truly additive, this model specification will lead to a more powerful test.  However, the additivity assumption is often used only for testing the location parameters in genetic association studies.

The expression of $E(d_i )=\sigma_i \sqrt{\frac{2}{\pi} (1-h_{ii})}$ in Section 2.2 suggests that the stage 2 regression of (\ref{di.eq}) could be improved by rescaling the $d_i$'s by $(1-h_{ii})^{-1/2}$.  This adjustment has been shown to improve the type 1 error control of Levene's original test for small samples with group design imbalance \citep{RN87} (independent observations with no group uncertainty implies  $h_{ii}=1/n_j$, where $n_j$ is the sample size of the group to which the $i$th observation belongs).  Examination of this rescaling for $gS$ under simulations involving correlated data, however, led to instances of increased type 1 error (results not shown).  Thus, further investigation is required to propose an appropriate adjustment.  Another potential improvement to the analysis of regression model (\ref{di.eq}) is from the recognition that the $d_i$'s are folded normals and are in fact slightly correlated through correlation between the estimated residuals, $\widehat {\varepsilon}_i$'s, even when there is no sample correlation among the true disturbances, $\varepsilon_i$'s. \cite{RN211} derived expressions for the covariance matrix of $\bm{d}$ for independent observations with no group uncertainty, showing that the correlation across the $d_i$'s disappears as the group sizes increase.  For the complex data scenarios considered here, $gS_{LAD}$ appears robust for even small samples. Nevertheless, the potential for gain in efficiency by accounting for this type of correlation merits additional consideration.
%remains an open question. 

The developments here did not consider additional covariates, $\bm{z}$, e.g. age and sex in genetic association studies.  The extension is straightforward if the effects of $\bm{z}$ on $y$ are strictly on the mean.  In that case, including $\bm{z}$ as part of the design matrix in the stage 1 regression of (\ref{yi.eq}) suffices.  However, if $\bm{z}$ also influences the variance of $y$, not including $\bm{z}$ as part of the design matrix in the stage 2 regression of (\ref{di.eq}) may lead to increased type 1 error of testing the $\gamma_j$'s that are associated with the primary covariates of interest. This is the same phenomenon observed in location-testing where omitting potential confounders can lead to spurious association. 

Joint location-scale testing is becoming a popular method for complex outcome-covariate association data, where the conventional location-only analyses may be underpowered.  This scenario has received attention in many fields ranging from economics to climate dynamics \cite{RN80}, in addition to our motivating example of genetic epidemiology \citep{RN212}.  The proposed $gS$ test allows investigators to combine evidence from scale tests with existing generalized location tests via the $JLS$ testing framework of \citep{RN212}, previously proposed for 
independent samples without group membership uncertainty.  The CF application study showed that individual location or scale tests can provide more significant results when utilizing related individuals, which in turn may lead to a more powerful $gJLS$ test.

%Summary: the  original Levene's method $Lev$ is not adequate in the presence of sample correlation.  The LAD approach should be used in the stage 1 regression to obtain group-{\it medium}-adjusted residuals for a more robust stage 2 regression analysis.  When group sizes are unbalanced and relatively small, the $TW_{LAD}$ method of Iachine, et al. [2010] has increased type 1 error while the proposed $gS_{LAD}$ test is accurate.  Finally, in the presence of group uncertainty, incorporating the probabilistic data in the model increase power as compared to the ``best-guess" approach. 

%  If your paper refers to supplementary web material, then you MUST
%  include this section!!  See Instructions for Authors at the journal
%  website http://www.biometrics.tibs.org

\section{Supplementary Materials}

Web Appendix Sections A, B and C, Web Figure 1, Web Tables 1-4, and R-code description for data analysis referenced in Sections 2, 3, 4, and 5 are available below in this document.

\backmatter

%  The \backmatter command formats the subsequent headings so that they
%  are in the journal style.  Please keep this command in your document
%  in this position, right after the final section of the main part of 
%  the paper and right before the Acknowledgements, Supplementary Materials,
%  and References sections. 

%  This section is optional.  Here is where you will want to cite
%  grants, people who helped with the paper, etc.  But keep it short!

\section*{Acknowledgements}

The authors thank Professor Jerry F. Lawless and Dr. Lisa J. Strug for helpful suggestions and
critical reading of the original version of the paper.  The authors thank Dr. Andrew Paterson and Dr. Lisa J. Strug for providing the type 1 diabetes and the cystic fibrosis application data, respectively.
This research is funded by the Natural Sciences and Engineering Research Council of Canada (NSERC 250053-2013 to LS) and the Canadian Institutes of Health Research (CIHR 201309MOP-117978 to LS).  DS is a trainee of the CIHR STAGE (Strategic Training in Advanced Genetic Epidemiology) training program at the University of Toronto and is a recipient of the SickKids Restracomp Studentship Award and the Ontario Graduate Scholarship (OGS).
%and an anonymous referee for very useful comments that improved the presentation of the paper.\vspace*{-8pt}

%  Here, we create the bibliographic entries manually, following the
%  journal style.  If you use this method or use natbib, PLEASE PAY
%  CAREFUL ATTENTION TO THE BIBLIOGRAPHIC STYLE IN A RECENT ISSUE OF
%  THE JOURNAL AND FOLLOW IT!  Failure to follow stylistic conventions
%  just lengthens the time spend copyediting your paper and hence its
%  position in the publication queue should it be accepted.

%  We greatly prefer that you incorporate the references for your
%  article into the body of the article as we have done here 
%  (you can use natbib or not as you choose) than use BiBTeX,
%  so that your article is self-contained in one file.
%  If you do use BiBTeX, please use the .bst file that comes with 
%  the distribution.

\bibliographystyle{biom} \bibliography{DraftMay18_2016_ref}

\begin{table}%[ht]
\small
\centering
\vspace*{1 cm}

\caption{\textbf{Type 1 error evaluation under simulation model 1.}  Six different tests were evaluated, including the original Levene's test, $Lev$, the twin test of \protect\cite{RN182}, $TW$, and the proposed generalized scale test, $gS$, with subscripts $_{OLS}$ and $_{LAD}$ denoting whether the stage 1 regression was performed using OLS or LAD. Parameter values included $n_1$ and $n_2$ for the number of MZ and DZ twin pairs, respectively, and $\rho_1=0.75$ and $\rho_2=0.5$ for the corresponding within-pair correlations.  Without loss of generality, $\sigma_1^2=\sigma_2^2=1$  for type 1 error rate evaluation.  The empirical type 1 error was estimated from 10,000 simulated replicates at the nominal 5\% level.
%The nominal $\alpha$ level is $5\%$ and the empirical value was estimated using 10,000 replicates.
\label{table1}}
\bigskip

\begin{tabular}{rrrrrrrr}
  \hline
$n_1$ & $n_2$ & $Lev_{OLS}$ & $Lev_{LAD}$ & $TW_{OLS}$ & $TW_{LAD}$ & $gS_{OLS}$ & $gS_{LAD}$  \\ 
  \hline
  \noalign{\vskip 2mm}
   &    &      \multicolumn{6}{c}{Gaussian}  \\  
    \noalign{\vskip 1mm}
20 & 20 & 0.102 & 0.087 & 0.055 & 0.044 & 0.058 & 0.046 \\ 
  5 & 5 & 0.115 & 0.071 & 0.085 & 0.041 & 0.099 & 0.049 \\ 
  10 & 20 & 0.112 & 0.091 & 0.085 & 0.064 & 0.075 & 0.054 \\ 
  5 & 10 & 0.114 & 0.079 & 0.118 & 0.079 & 0.092 & 0.054 \\ 
    \noalign{\vskip 2mm}
  &     &      \multicolumn{6}{c}{Student's $t_4$}  \\  
        \noalign{\vskip 1mm}
  20 & 20 & 0.102 & 0.084 & 0.056 & 0.043 & 0.059 & 0.045 \\ 
  5 & 5 & 0.129 & 0.069 & 0.086 & 0.037 & 0.103 & 0.046 \\ 
  10 & 20 & 0.118 & 0.093 & 0.090 & 0.069 & 0.078 & 0.054 \\ 
  5 & 10 & 0.123 & 0.076 & 0.115 & 0.071 & 0.093 & 0.048 \\ 
     \noalign{\vskip 2mm}
     &     &      \multicolumn{6}{c}{$\chi_4^2$}  \\  
            \noalign{\vskip 1mm}
  20 & 20 & 0.175 & 0.098 & 0.112 & 0.052 & 0.117 & 0.054 \\ 
  5 & 5 & 0.180 & 0.083 & 0.133 & 0.053 & 0.153 & 0.061 \\ 
  10 & 20 & 0.181 & 0.102 & 0.146 & 0.079 & 0.137 & 0.062 \\ 
  5 & 10 & 0.187 & 0.094 & 0.178 & 0.085 & 0.149 & 0.064 \\ 
     \hline
\end{tabular}
\end{table}

\begin{table}%[htb]
\small
\centering
\vspace*{1 cm}

\caption{\textbf{Type 1 error evaluation under simulation model 2 without group uncertainty.}  
Parameter values included $n/2$ for the number of sib-pairs, $\rho=0.5$ for the within-pair correlations, and $q$ = 0.1 or 0.2 for the minor allele frequency (MAF) of the SNP of interest; on average the expected sizes of the three genotype groups are $n(1-q)^2$, $n2q(1-q)$ and $nq^2$.  Without loss of generality, $\sigma_0^2=\sigma_1^2=\sigma_2^2=1$ for type 1 error rate evaluation.  
The empirical type 1 error was estimated from 10,000 simulated replicates at the nominal 5\% level.
\label{table2}}
\begin{tabular}{rrrrrrr}
  \Hline
     \noalign{\vskip 1mm}
    $n/2$ &      \multicolumn{2}{c}{Gaussian}   &      \multicolumn{2}{c}{Student's $t_4$}  &      \multicolumn{2}{c}{$\chi_4^2$} \\  
    \cmidrule(r){1-1} \cmidrule(lr){2-3} \cmidrule(lr){4-5}  \cmidrule(lr){6-7} 

   & $TW_{LAD}$ & $gS_{LAD}$ &$TW_{LAD}$ & $gS_{LAD}$  &$TW_{LAD}$ & $gS_{LAD}$ \\ 
 \cmidrule(lr){2-3} \cmidrule(lr){4-5}  \cmidrule(lr){6-7} 
   \noalign{\vskip 2mm}

   \noalign{\vskip 1mm}
     &      \multicolumn{6}{c}{MAF=0.1}  \\
20 & 0.110 & 0.040 & 0.109 & 0.042 & 0.113 & 0.044 \\ 
  50 & 0.117 & 0.043 & 0.140 & 0.046 & 0.160 & 0.044 \\ 
  100 & 0.092 & 0.048 & 0.115 & 0.049 & 0.118 & 0.047 \\ 
  500 & 0.056 & 0.048 & 0.068 & 0.047 & 0.070 & 0.052 \\ 
  1000 & 0.055 & 0.050 & 0.061 & 0.049 & 0.058 & 0.045 \\ 
   \noalign{\vskip 1mm}
     &      \multicolumn{6}{c}{MAF=0.2}  \\
  20 & 0.068 & 0.039 & 0.072 & 0.040 & 0.092 & 0.050 \\ 
  50 & 0.074 & 0.042 & 0.086 & 0.041 & 0.095 & 0.046 \\ 
  100 & 0.060 & 0.048 & 0.072 & 0.044 & 0.078 & 0.051 \\ 
  500 & 0.055 & 0.051 & 0.055 & 0.047 & 0.057 & 0.052 \\ 
  1000 & 0.051 & 0.051 & 0.053 & 0.051 & 0.056 & 0.051 \\ 
      \hline
\end{tabular}
\end{table}

\begin{table}%[ht]
\small
\centering
\vspace*{1 cm}

\caption{\textbf{Type 1 error evaluation under simulation model 2 with 30\% group uncertainty.}  Superscript $^{BG}$ denotes $TW_{LAD}$ and  $gS_{LAD}$ applied to the ``best-guess" genotype data. The true genotype data were masked using a Dirichlet distribution for the genotype probabilities with scale parameters $a$ for the correct genotype and $(1-a)/2$ for the other two.  On average, $a=0.7$ corresponds to 30\% group uncertainty level.  See legend of Table \ref{table2} for additional simulation details. 
\label{table3}}
\begin{adjustbox}{center}

\begin{tabular}{rrrrrrrrrrrrr}
  \Hline
     \noalign{\vskip 1mm}
    $n/2$ &      \multicolumn{4}{c}{Gaussian}   &      \multicolumn{4}{c}{Student's $t_4$}  &      \multicolumn{4}{c}{$\chi_4^2$} \\  
    \cmidrule(r){1-1}  \cmidrule(lr){2-5} \cmidrule(lr){6-9}  \cmidrule(lr){10-13} 

 & $TW_{LAD}^{BG}$ & $TW_{LAD}$  & $gS_{LAD}^{BG}$ & $gS_{LAD}$ & $TW_{LAD}^{BG}$ & $TW_{LAD}$ &$gS_{LAD}^{BG}$ & $gS_{LAD}$ & $TW_{LAD}^{BG}$ & $TW_{LAD}$ &$gS_{LAD}^{BG}$ & $gS_{LAD}$ \\ 
 \cmidrule(lr){2-3} \cmidrule(lr){4-5}  \cmidrule(lr){6-7}  \cmidrule(lr){8-9} \cmidrule(lr){10-11}  \cmidrule(lr){12-13} 
   \noalign{\vskip 2mm}

   \noalign{\vskip 1mm}
     &      \multicolumn{12}{c}{MAF=0.1}  \\
20 & 0.067 & 0.074 & 0.036 & 0.037 & 0.083 & 0.079 & 0.044 & 0.046 & 0.088 & 0.090 & 0.047 & 0.050 \\ 
  50 & 0.066 & 0.062 & 0.045 & 0.045 & 0.076 & 0.062 & 0.046 & 0.046 & 0.084 & 0.076 & 0.049 & 0.053 \\ 
  100 & 0.058 & 0.057 & 0.045 & 0.046 & 0.064 & 0.059 & 0.047 & 0.046 & 0.072 & 0.069 & 0.051 & 0.049 \\ 
  500 & 0.057 & 0.054 & 0.054 & 0.052 & 0.056 & 0.053 & 0.052 & 0.048 & 0.055 & 0.052 & 0.050 & 0.048 \\ 
  1000 & 0.051 & 0.055 & 0.052 & 0.054 & 0.053 & 0.050 & 0.052 & 0.049 & 0.054 & 0.052 & 0.049 & 0.047 \\ 
     \noalign{\vskip 1mm}
     &      \multicolumn{12}{c}{MAF=0.2}  \\
  20 & 0.061 & 0.062 & 0.040 & 0.039 & 0.065 & 0.063 & 0.037 & 0.044 & 0.075 & 0.073 & 0.047 & 0.050 \\ 
  50 & 0.053 & 0.053 & 0.046 & 0.045 & 0.063 & 0.059 & 0.046 & 0.050 & 0.069 & 0.070 & 0.051 & 0.053 \\ 
  100 & 0.049 & 0.051 & 0.046 & 0.045 & 0.057 & 0.053 & 0.047 & 0.049 & 0.059 & 0.058 & 0.049 & 0.047 \\ 
  500 & 0.051 & 0.049 & 0.049 & 0.051 & 0.051 & 0.052 & 0.048 & 0.050 & 0.053 & 0.052 & 0.048 & 0.051 \\ 
  1000 & 0.049 & 0.046 & 0.047 & 0.047 & 0.052 & 0.049 & 0.047 & 0.049 & 0.055 & 0.053 & 0.050 & 0.052 \\ 
      \hline
\end{tabular}
\end{adjustbox}

\end{table}

\begin{table}%[ht]
\small
\centering
\vspace*{1 cm}

\caption{\textbf{Power of $gS_{LAD}^{BG}$ and $gS_{LAD}$ under simulation model 2 with 30\% group uncertainty.}
$gS_{LAD}^{BG}$ denotes $gS_{LAD}$ applied to the ``best-guess" genotype data.  The true genotypes were masked using a Dirichlet distribution for the genotype probabilities with scale parameters $a$ for the correct genotype and $(1-a)/2$ for the other two. On average, $a=0.7$ corresponds to 30\% group uncertainty. Besides the parameters shown in the table, other values include  $\rho=0.5$ for within-pair correlation, and 
$(\sigma_0^2,\sigma_1^2,\sigma_2^2)=(1, 1.5, 2)$. Power was estimated from 1,000 simulated replicates at the 5\% level.  
\label{table4}}

\begin{tabular}{rrrrrrr}
  \Hline
     \noalign{\vskip 1mm}
    $n/2$ &      \multicolumn{2}{c}{Gaussian}   &      \multicolumn{2}{c}{Student's $t_4$}  &      \multicolumn{2}{c}{$\chi_4^2$} \\  
    \cmidrule(r){1-1} \cmidrule(lr){2-3} \cmidrule(lr){4-5}  \cmidrule(lr){6-7} 

   & $gS_{LAD}^{BG}$ & $gS_{LAD}$ &$gS_{LAD}^{BG}$ & $gS_{LAD}$  &$gS_{LAD}^{BG}$ & $gS_{LAD}$ \\ 
 \cmidrule(lr){2-3} \cmidrule(lr){4-5}  \cmidrule(lr){6-7} 
   \noalign{\vskip 2mm}

   \noalign{\vskip 1mm}
     &      \multicolumn{6}{c}{MAF=0.1}  \\
20 & 0.064 & 0.066 & 0.067 & 0.077 & 0.050 & 0.064 \\ 
  50 & 0.079 & 0.087 & 0.077 & 0.081 & 0.089 & 0.089 \\ 
  100 & 0.124 & 0.152 & 0.087 & 0.112 & 0.101 & 0.117 \\ 
  500 & 0.495 & 0.613 & 0.314 & 0.420 & 0.376 & 0.442 \\ 
  1000 & 0.795 & 0.885 & 0.533 & 0.671 & 0.634 & 0.759 \\ 
   \noalign{\vskip 1mm}
     &      \multicolumn{6}{c}{MAF=0.2}  \\
20 & 0.050 & 0.066 & 0.062 & 0.058 & 0.063 & 0.074 \\ 
  50 & 0.089 & 0.120 & 0.084 & 0.089 & 0.091 & 0.104 \\ 
  100 & 0.166 & 0.196 & 0.114 & 0.129 & 0.129 & 0.160 \\ 
  500 & 0.668 & 0.784 & 0.471 & 0.582 & 0.499 & 0.608 \\ 
  1000 & 0.939 & 0.985 & 0.739 & 0.846 & 0.810 & 0.896 \\ 
      \hline
\end{tabular}
\end{table}

\begin{table}%[ht]
\small
\centering
\vspace*{1 cm}

\caption{\textbf{Application study of lung function in patients with cystic fibrosis.}  There were 1313 singletons, 94 sib-pairs and  2 sib-trios in the whole sample, resulting in $n_{indep}=1313+94+2=1409$ unrelated individuals, and   
$n_{all}=1313+94*2+2*3=1507$ individuals.  Results for $n_{indep}$ were from \protect\cite{RN212}, where the standard regression Location test, Levene's scale test and the $JLS$ joint location-scale test were used. Results for 
$n_{all}$ were obtained from the corresponding generalized tests, with LAD  used for the stage-1 regression for the $gS$ test.
\label{table5}}

%\vspace*{0.5cm}
\begin{tabular}{lllllrrrrrr}
  \Hline
  && &&& \multicolumn{3}{c}{\pbox{3.3cm}{$n_{indep}=1409$}}& \multicolumn{3}{c}{\pbox{3.3cm}{$n_{all}=1507$}}  \\ 
     \noalign{\vskip 1mm}
 Chr & Gene & SNP & bp-Position & MAF & $L$ocation & $S$cale & $JLS$ & $gL$ & $gS$ & $gJLS$ \\ 
\cmidrule(r){1-5} \cmidrule(lr){6-8} \cmidrule(l){9-11}
   \noalign{\vskip 2mm}
1&	\textit{SLC26A9} &	rs7512462 & 204,166,218 & 0.41 & 0.30 & 0.58 & 0.48 & 0.30 & 0.39 & 0.36 \\ 
1&	\textit{SLC26A9} &	rs4077468 & 204,181,380 & 0.42 & 0.53 & 0.61 & 0.69 & 0.45 & 0.59 & 0.62 \\ 
1&	\textit{SLC26A9} &	rs12047830 & 204,183,322 & 0.49 & 0.55 & 0.15 & 0.29 &	 0.52 & 0.11 & 0.22 \\ 
1&	\textit{SLC26A9} &	rs7419153 & 204,183,932 & 0.37 & 0.50 & 0.06 & 0.14 & 0.73 & 0.09 & 0.24 \\ 
5&	\textit{SLC9A3} &	rs17563161 & 550,624 & 0.26 & 0.0004 & 0.02 & 0.0001 & 0.0002 & 0.02 & 5.6x10$^{-5}$ \\ 
X&	\textit{SLC6A14} &	rs12839137 & 115,479,578 & 0.24 & 0.02 & 0.08 & 0.01 & 0.01 & 0.16 & 0.02 \\ 
X&	\textit{SLC6A14}  &	rs5905283 & 115,479,909 & 0.49 & 0.009 & 0.07 & 0.005 & 0.005 & 0.18 & 0.007\\ 
X&	\textit{SLC6A14}  &	rs3788766 & 115,480,867 & 0.40 & 0.001 & 0.01 & 0.0002 & 0.0004 & 0.02 & 9.5x10${-5}$\\ 
      \hline
\end{tabular}
\end{table}

\label{lastpage}

%Include the supplemental pdf here!
 \afterpage{

\section*{Supplementary material}

# Web-based Supplementary Materials for "A Generalized Levene's Scale Test for Variance Heterogeneity in the Presence of Sample Correlation and Group Uncertainty"

by David Soave and Lei Sun

**Web Appendix A: Asymptotic Distribution of $gS$**

A number of authors have investigated the asymptotic distribution of test statistics based on absolute residuals from independent observations with respect to estimation effects; effects on stage 2 regression analysis when using absolute residuals, $d_i = |\widehat{\varepsilon}_i| = \left|y_i - \boldsymbol{x_i}^T\widehat{\boldsymbol{\beta}}\right|$, from stage 1 instead of absolute true disturbances $|\varepsilon_i| = \left|y_i - \boldsymbol{x_i}^T\boldsymbol{\beta}\right|$ (Godfrey, 1996; Im, 2000; Machado and Silva, 2000; Furno, 2005; Carroll and Schneider, 1985). In general, stage 2 tests involving $d_i$ from stage 1 OLS regression are only asymptotically correct when the $y_i$ follow a symmetric distribution. Conversely, stage 2 tests involving $d_i$ from stage 1 LAD regression will be asymptotically correct regardless of the shape of the error distribution (Furno, 2005; Carroll and Schneider, 1985). We demonstrate here that these results hold for the proposed $gS$ test when using the $F$-statistic in the stage 2 generalized least squares regression to account for possible data dependence due to clusters/family relationships.

In keeping with the notation from the main text, we use only a single subscript (e.g. $i = 1, \ldots, n$) to represent individuals across the sample and rely on the covariance matrix, $\Sigma_{stage\ 2} = \sigma_d^2 \Sigma_d$, to specify relationships between family/cluster members. Note that we are assuming homoscedasticity across the $d_i$'s and thus $\Sigma_d = CC^T$ describes the Cholesky decomposition of the correlation across $\boldsymbol{d}$. Let $f_i = \left|y_i - \boldsymbol{x_i}^T\boldsymbol{\beta}\right|$, represent the absolute true disturbances from the stage 1 model (2), where $\boldsymbol{\beta}$ represents the true central tendency parameter vector being estimated (i.e. mean or median). For fixed $C$, let $\boldsymbol{f}^* = C^{-1}\boldsymbol{f}$, such that for the model $\boldsymbol{f}^* = \alpha \boldsymbol{1}^* + Z^*\boldsymbol{\gamma} + \boldsymbol{e}^*$, $E(\boldsymbol{e}^*) = \boldsymbol{0}$ and $Var(\boldsymbol{e}^*) = \sigma_f^2 I$. Note that $\boldsymbol{1}^* = C^{-1}\boldsymbol{1}$

and $Z^* = C^{-1}Z$ are a partitioned transformation of the $gS$ stage 1 regression design matrix (i.e. $X = (\mathbf{1}, Z)$ from equation (3)). Also note that the $f$ subscript of $\sigma_f^2$ defines the variance parameter associated with $\boldsymbol{f}$ conditional on $C$ and $X$. We assume that we are able estimate $\boldsymbol{\beta}$, and that $n^{1/2}(\widehat{\boldsymbol{\beta}} - \boldsymbol{\beta})$ has an asymptotic multivariate normal distribution with finite covariance matrix; this would be true for either OLS or LAD stage 1 regression estimates of $\boldsymbol{\beta}$.

We consider an $F$-test based on $d_i^*$, which are the elements of $\boldsymbol{d}^* = \boldsymbol{C}^{-1}\boldsymbol{d}$,

$$F(\boldsymbol{d}^*) = \frac{MSR(\boldsymbol{d}^*)}{MSE(\boldsymbol{d}^*)} = \frac{\sum_i^n (\widehat{d_i^*} - \widetilde{d_i^*})^2/(k-1)}{\sum_i^n (d_i^* - \widehat{d_i^*})^2/(n-k)},$$

where $\widehat{d_i^*} = (\boldsymbol{x}_i^*)^T\widehat{\boldsymbol{\theta}}$, the predicted values from the regression equation (7), and $\widetilde{d_i^*} = 1_i^*\widetilde{\alpha}$, the predicted values from the regression of $\boldsymbol{d}^*$ on $\mathbf{1}^*$.

We first note that, in the absence of estimation effects, $(k-1)F(\boldsymbol{f}^*)$ is asymptotically distributed as chi-squared with $k-1$ degrees of freedom (see Arnold (1980)). To obtain the asymptotic distribution of $F(\boldsymbol{d}^*)$ we need to assume that $(\boldsymbol{x}_i, \varepsilon_i)$ satisfy regularity conditions necessary for the application of standard asymptotic theory. Specifically, we assume

ASSUMPTION 1:   $E(\boldsymbol{x}_i \varepsilon_i) = \mathbf{0}$

ASSUMPTION 2:   $E(\boldsymbol{x}_i \boldsymbol{x}_i^T)$ is positive definite

ASSUMPTION 3:   $(y_i, \boldsymbol{x}_i^T)$ has finite fourth moment.

Similar to the assertion of Carroll and Schneider (1985), we have that $MSE(\boldsymbol{d}^*) \xrightarrow{p} \sigma^2 = Var(f_i^*)$.

We can also rewrite the numerator of $F_n(\boldsymbol{d}^*)$ as

$$MSR(\boldsymbol{d}^*) = \frac{1}{k-1} \sum_{i=1}^n d_i^*(\boldsymbol{z}_i^* - \widetilde{\boldsymbol{z}_i^*})^T \left( \sum_{i=1}^n (\boldsymbol{z}_i^* - \widetilde{\boldsymbol{z}_i^*})(\boldsymbol{z}_i^* - \widetilde{\boldsymbol{z}_i^*})^T \right)^{-1} \sum_{i=1}^n (\boldsymbol{z}_i^* - \widetilde{\boldsymbol{z}_i^*})d_i^*.$$

where $(\boldsymbol{z}_i^*)^T$ are the row vectors of $Z^*$, and $\widetilde{\boldsymbol{z}_i^*}$ are the predicted values from regressing the column(s) of $Z^*$ on $\mathbf{1}^*$. Note that $\mathbf{1}^*$ will not, in general, be a vector or 1's, and will only be

constant when all observations come from equal sized clusters with an identical symmetric correlation structure.

Let $t = 1, ..., T$ and $s = 1, ..., S$ index the variables associated with the residuals, $\widehat{\varepsilon}_i$, that take positive and negative values, respectively. Correspondingly, $T$ and $S$ are the number of residuals that take positive and negative values, respectively. Let $c_{i,j}$ represent the element for the $i$th row and $j$th column of the matrix $C^{-1}$, and let $[-]_i$ denote the $i$th element from the vector with the square brackets.

Finally, we make the additional assumption,

ASSUMPTION 4: $\quad E(\bm{x}_i \bm{x}_i^T | \varepsilon_i > 0) = E(\bm{x}_i \bm{x}_i^T | \varepsilon_i < 0).$

This assumption follows from Assumption 1, that $\bm{x}_i$ and $\varepsilon_i$ are uncorrelated, and implies

$$\frac{1}{T} \sum_{i=1}^{n} \sum_{t=1}^{T} (\bm{z}_i^* - \widetilde{\bm{z}}_i^*) c_{i,t} \bm{x}_t^T = \frac{1}{S} \sum_{i=1}^{n} \sum_{s=1}^{S} (\bm{z}_i^* - \widetilde{\bm{z}}_i^*) c_{i,s} \bm{x}_s^T + o_p(1)$$
$$= \frac{1}{n} \sum_{i=1}^{n} \sum_{j=1}^{n} (\bm{z}_i^* - \widetilde{\bm{z}}_i^*) c_{i,j} \bm{x}_j^T + o_p(1).$$

Now, applying the results of Lemma 1 from Im (2000), we propose the following lemma with proof similar to Im (2000) Lemma 3,

**Lemma 1**. Under Assumptions 1-4,

$$\frac{1}{\sqrt{n}} \sum_{i=1}^{n} (\bm{z}_i^* - \widetilde{\bm{z}}_i^*) d_i^* - \frac{1}{\sqrt{n}} \sum_{i=1}^{n} (\bm{z}_i^* - \widetilde{\bm{z}}_i^*) f_i^*$$
$$= -(2Pr(\varepsilon_l > 0) - 1) \frac{1}{n} \sum_{i=1}^{n} \sum_{j=1}^{n} (\bm{z}_i^* - \widetilde{\bm{z}}_i^*) c_{i,j} \bm{x}_j^T \sqrt{n}(\widehat{\bm{\beta}} - \bm{\beta}) + o_p(1).$$

*Proof.*

LHS

$$= \frac{1}{\sqrt{n}} \sum_{i=1}^{n} (z_i^* - \widetilde{z}_i^*)(d_i^* - f_i^*)$$

$$= \frac{1}{\sqrt{n}} \sum_{i=1}^{n} (z_i^* - \widetilde{z}_i^*) \left( \left[C^{-1}d\right]_i - \left[C^{-1}f\right]_i \right) + o_p(1)$$

$$= \frac{1}{\sqrt{n}} \sum_{i=1}^{n} (z_i^* - \widetilde{z}_i^*) \left( -\sum_{t=1}^{T} c_{i,t} x_t^T (\widehat{\beta} - \beta) + \sum_{s=1}^{S} c_{i,s} x_s^T (\widehat{\beta} - \beta) \right) + o_p(1)$$

$$= \frac{1}{\sqrt{n}} \sum_{i=1}^{n} (z_i^* - \widetilde{z}_i^*) \left( -\frac{T}{n} \frac{1}{T} \sum_{t=1}^{T} c_{i,t} x_t^T + \frac{S}{n} \frac{1}{S} \sum_{s=1}^{S} c_{i,s} x_s^T \right) (\widehat{\beta} - \beta) + o_p(1)$$

$$= \sum_{i=1}^{n} (z_i^* - \widetilde{z}_i^*) \left( -\frac{T}{n} \frac{1}{T} \sum_{t=1}^{T} c_{i,t} x_t^T + \frac{S}{n} \frac{1}{S} \sum_{s=1}^{S} c_{i,s} x_s^T \right) \sqrt{n}(\widehat{\beta} - \beta) + o_p(1)$$

$$= \left( -\frac{T}{n} \frac{1}{T} \sum_{i=1}^{n} \sum_{t=1}^{T} (z_i^* - \widetilde{z}_i^*) c_{i,t} x_t^T + \frac{S}{n} \frac{1}{S} \sum_{i=1}^{n} \sum_{s=1}^{S} (z_i^* - \widetilde{z}_i^*)^T c_{i,s} x_s^T \right) \sqrt{n}(\widehat{\beta} - \beta) + o_p(1)$$

$$= -\frac{(T-S)}{n} \frac{1}{n} \sum_{i=1}^{n} \sum_{j=1}^{n} (z_i^* - \widetilde{z}_i^*) c_{i,j} x_j^T \sqrt{n}(\widehat{\beta} - \beta) + o_p(1) \quad \blacksquare$$

If the final equation becomes negligible as $n$ becomes large, we can say that the absolute residuals, $d$, may replace the true disturbances, $f$, where

$$MSR(d^*) - MSR(f^*) \xrightarrow{p} 0,$$

and consequently

$$F(d^*) - F(f^*) \xrightarrow{p} 0.$$

This, in general, will only be the case when $(T - S)/n = o_p(1)$, which will occur for stage 1 OLS regression only if the error distribution is symmetric. On the other hand, $(T - S)/n = o_p(1)$ is ensured for stage 1 LAD regression as minimization of the sum of absolute residuals must equate the number of positive and negative residuals (Koenker and Hallock, 2001). Note that Lemma 1 and its proof reduce to the same result of Im (2000) Lemma 3 for the case where $\Sigma_d = I$ (independent errors).

This leads to the following result.

**Theorem 1**. Under the previously stated conditions, if $(T - S)/n = o_p(1)$,

$$(k - 1)F(\boldsymbol{d}^*) \xrightarrow{d} \chi^2(k - 1).$$

## Web Appendix B: Independence between the generalized location ($gL$) and scale ($gS$) test statistics

To test for a location (mean) effect of a set of $(k - 1)$ covariates, $\boldsymbol{x}_i^T = (1, x_{i1}, \ldots, x_{i(k-1)})$ including an intercept, on a quantitative outcome $y_i$ with known correlation among the $y_i$'s, we can obtain the $F$-statistic from a generalized least squares regression of the linear model,

$$\boldsymbol{y} = X\boldsymbol{\beta} + \boldsymbol{\varepsilon}, \quad \boldsymbol{\varepsilon} \sim \mathcal{N}(\boldsymbol{0}, \Sigma_{stage1}), \tag{1-1}$$

where the design matrix, $X$, represents the stacking of the $\boldsymbol{x}_i^T$'s, and $\Sigma_{stage1}$ represents the covariance matrix for $\boldsymbol{y}$. We include indexes for families $(i = 1, \ldots, M)$ and members within families $(j = 1, \ldots, n_i)$ to specify models that allows for within group/family dependence of observations as would be encountered in sibling or pedigree data. Under the assumption of homoscedasticity, $\Sigma_{stage1}$ can be written as $\sigma_y^2 \Sigma_y$, where the correlation matrix, $\Sigma_y$, specifies relationships between family/cluster members.

Let $\Sigma_y = C_y C_y^T$ be the Cholesky decomposition, and for fixed $\Sigma_y$, we use the following transformations,

$$\boldsymbol{y}^* = C^{-1}\boldsymbol{y}, \quad X^* = C^{-1}X, \quad \boldsymbol{\varepsilon}^* = C^{-1}\boldsymbol{\epsilon},$$

Thus, we can rewrite (1-1) as a classical linear model,

$$\boldsymbol{y}^* = X^*\boldsymbol{\beta} + \boldsymbol{\varepsilon}^*, \quad \boldsymbol{\varepsilon}^* \sim \mathcal{N}(\boldsymbol{0}, \sigma_y^2 I).$$

Thus, the generalized location ($gL$) test statistic ($T_{Location}$) is simply the $F$-statistic from an ordinary least squares regression of $\boldsymbol{y}^*$ on $X^*$, which can be written as

$$T_{Location} = F(\boldsymbol{y}^*) = \frac{MSR(\boldsymbol{y}^*)}{MSE(\boldsymbol{y}^*)} = \frac{\sum_i^n (\widehat{y_i^*} - \widetilde{y_i^*})^2/(k-1)}{\sum_i^n (y_i^* - \widehat{y_i^*})^2/(n-k)},$$

where $\widetilde{y_i^*} = 1_i^* \widetilde{\beta_0}$ and $\widehat{y_i^*} = \boldsymbol{x}_i^{*T} \widehat{\boldsymbol{\beta}}_{all}$, are the predicted values from the regressions of $\boldsymbol{y}^*$ on the first column of $X^*$(yielding $\widetilde{\beta}_0$) and all columns of $X^*$(yielding $\widehat{\boldsymbol{\beta}}_{all}$), respectively.

Without loss of generality, we assume the analysis of genetic data involving related subjects (family data) with an additive genetic model where $y_i$ is the quantitative trait under investigation and $\boldsymbol{x}_i^T = (1, x_i)$, where $x_i = 0, 1$ or $2$, reflects the number of minor alleles at a genetic variant under analysis, or the genotype dosage estimate incorporating uncertainty. Thus, the density of $\boldsymbol{y}$ is an exponential family with three parameters $\boldsymbol{\theta} = (\theta_1, \theta_2, \theta_3) = (\frac{\beta_1}{\sigma^2}, -\frac{1}{2\sigma^2}, \frac{\beta_0}{\sigma^2})$, for which the sufficient statistics $\boldsymbol{T} = (T_1, T_2, T_3) = (\sum x_i^* y_i^*, \sum y_i^{*2}, \sum 1_i^* y_i^*)$ are complete. Note that $1_i^*$ and $x_i^*$ are obtained from $X^*$, and $y_i^*$ from $\boldsymbol{y}^*$ defined above.

Let $T_{Scale}$ be the proposed $gS$ test statistic (see Section 2.3, equation (8)) analyzing the same covariate vector specified for $T_{Location}$. Now we have the following Lemma.

**Lemma 2**: For the conditional normal model $\boldsymbol{y} \sim \mathcal{N}(X\boldsymbol{\beta}, \sigma_y^2 \Sigma_y)$ with fixed $\Sigma_y$, $T_{Location}$ and $T_{Scale}$ are independent.

Proof. We showed that if $\Sigma_{stage2} = \sigma_d^2 \Sigma_d$, with $\Sigma_d$ fixed, and the covariate design vector $\boldsymbol{x}_i^T$ is length $k$, including intercept, then $T_{Scale}$ is asymptotically distributed as chi-squared with $k-1$ degrees of freedom (Theorem 1, Web Appendix A), and it does not depend on $\boldsymbol{\theta}$ (i.e. $T_{Scale}$ is ancillary for $\boldsymbol{\theta}$). Thus, $T_{Scale}$ is independent of $\boldsymbol{T}$ (see page 152 in Lehmann and Romano (2005)). Because $T_{Location}$ is a function of $\boldsymbol{T}$, $T_{Location}$ and $T_{Scale}$ are therefore independent. ∎

### Web Appendix C: Implementing $gS$, $TW$, and $Lev$ tests

Under the simulation model 1 setup, we compared the performance of three tests: Levene's original test procedure (using two measures of central tendency: the sample means and sample medians, implemented in R using the "lawstat" package), the $TW$ test proposed

by Iachine et al. (2010) (using the same two measures of central tendency, with stage 2 regression implemented in R (see below) to mimic the results of the *regress*, *cluster*() command implemented in STATA), and our proposed $gS$ test (using the same two measures of central tendency, with stage 2 regression implemented in R using the *gls*() function in the "nlme" package).

For the different parameter combinations of simulation model 2, we considered two measures of central tendency from stage 1 of the procedure: the predicted values from OLS and LAD regression. All testing procedures were implemented in R as just described, while using the *rq*() function from the "quantreg" package to perform stage 1 LAD regression. [R Core Team (2016). R: A language and environment for statistical computing. R Foundation for Statistical Computing, Vienna, Austria. URL https://www.R-project.org/.]

*TW: Implementing the cluster-robust stage 2 regression in R*

Numerous modifications have been proposed for various types of cluster-designed data sets (Cameron and Miller, 2013). Iachine et al. (2010) chose to implement $TW$ with the STATA statistical software command *regress* using the *cluster*() option to incorporate the cluster information in a standard error adjustment. This method of cluster robust inference can be implemented with the R software as follows: First, fit the desired panel data model (e.g. $\boldsymbol{d} \sim X$ for the stage 2 regression; equation (4) of Section 2.2) using the *plm*() function from the "plm" package, with option *index*="cluster.id". Next, estimate the cluster robust covariance matrix using the *vcovHC*() function from the "sandwich" package, with options type= "$HC0$" and *adjust* $= T$. Next, multiply the covariance matrix by the following adjustment factor:

$$\frac{M}{M-1} \cdot \frac{n-1}{n-k},$$

where $M$ is the total number of clusters and $n-k$ is the residual degrees of freedom from the plm model. Next, perform the coefficient test using the *waldtest*() function from the

"lmtest" package, setting the *vcov* option equal to the new covariance matrix and the *test* option equal to "*F*". Finally, the resulting $F$-statistic is then compared to an $F$-distribution with ($k$-1) and ($M$-1) degrees of freedom to obtain the $p$-value of the $TW$ test. (Note that the default degrees of freedom for *waldtest*() will be ($k$-1) and ($N$-1).)

**Web Tables and Figure**

Table 1

***Type 1 error evaluation under simulation model 1 (large samples).*** *Six different tests were evaluated, including the original Levene's test, Lev, the twin test of Iachine et al. (2010), TW, and the proposed generalized scale test, gS, with subscripts $_{OLS}$ and $_{LAD}$ denoting whether the stage 1 regression was performed using OLS or LAD. Parameter values include $n_1$ and $n_2$ for the number of MZ and DZ twin pairs, respectively, and $\rho_1 = 0.75$ and $\rho_2 = 0.5$ for the corresponding within-pair correlations. Without loss of generality, $\sigma_1^2 = \sigma_2^2 = 1$ for type 1 error rate evaluation. The empirical type 1 error was estimated from 10,000 simulated replicates at the nominal 5% level.*

| $n_1$ | $n_2$ | $Lev_{OLS}$ | $Lev_{LAD}$ | $TW_{OLS}$ | $TW_{LAD}$ | $gS_{OLS}$ | $gS_{LAD}$ |
|---|---|---|---|---|---|---|---|
| | | | | Gaussian | | | |
| 2000 | 2000 | 0.097 | 0.097 | 0.050 | 0.050 | 0.050 | 0.050 |
| 500 | 500 | 0.094 | 0.093 | 0.049 | 0.048 | 0.049 | 0.048 |
| 1000 | 2000 | 0.102 | 0.102 | 0.050 | 0.050 | 0.058 | 0.058 |
| 500 | 1000 | 0.103 | 0.102 | 0.053 | 0.052 | 0.062 | 0.061 |
| | | | | Student's $t_4$ | | | |
| 2000 | 2000 | 0.091 | 0.091 | 0.045 | 0.045 | 0.045 | 0.045 |
| 500 | 500 | 0.094 | 0.093 | 0.048 | 0.047 | 0.048 | 0.047 |
| 1000 | 2000 | 0.097 | 0.098 | 0.046 | 0.046 | 0.055 | 0.055 |
| 500 | 1000 | 0.107 | 0.105 | 0.050 | 0.050 | 0.059 | 0.058 |
| | | | | $\chi_4^2$ | | | |
| 2000 | 2000 | 0.164 | 0.102 | 0.103 | 0.052 | 0.103 | 0.052 |
| 500 | 500 | 0.161 | 0.100 | 0.103 | 0.047 | 0.103 | 0.047 |
| 1000 | 2000 | 0.179 | 0.111 | 0.109 | 0.052 | 0.123 | 0.060 |
| 500 | 1000 | 0.173 | 0.107 | 0.104 | 0.052 | 0.116 | 0.060 |

**Table 2**

***Type 1 error evaluation under simulation model 2 with 10% group uncertainty.*** *Superscript* $^{BG}$ *denotes* $TW_{LAD}$ *and* $gS_{LAD}$ *being applied to the "best-guess" genotype data. The true genotype data were masked using a Dirichlet distribution for the genotype probabilities with scale parameters a for the correct genotype and* $(1-a)/2$ *for the other two. On average* $a = 0.9$ *corresponds to 10% group uncertainty. See legend of Table 2 (main text) for additional simulation details.*

| $n/2$ | Gaussian | | | | Student's $t_4$ | | | | $\chi^2_4$ | | | |
|---|---|---|---|---|---|---|---|---|---|---|---|---|
| | $TW^{BG}_{LAD}$ | $TW_{LAD}$ | $gS^{BG}_{LAD}$ | $gS_{LAD}$ | $TW^{BG}_{LAD}$ | $TW_{LAD}$ | $gS^{BG}_{LAD}$ | $gS_{LAD}$ | $TW^{BG}_{LAD}$ | $TW_{LAD}$ | $gS^{BG}_{LAD}$ | $gS_{LAD}$ |
| | | | | | MAF=0.1 | | | | | | | |
| 20 | 0.089 | 0.165 | 0.041 | 0.034 | 0.097 | 0.165 | 0.044 | 0.039 | 0.106 | 0.182 | 0.053 | 0.049 |
| 50 | 0.097 | 0.088 | 0.043 | 0.041 | 0.114 | 0.108 | 0.046 | 0.044 | 0.126 | 0.122 | 0.046 | 0.047 |
| 100 | 0.071 | 0.071 | 0.045 | 0.047 | 0.088 | 0.084 | 0.046 | 0.046 | 0.102 | 0.099 | 0.050 | 0.047 |
| 500 | 0.056 | 0.056 | 0.050 | 0.052 | 0.061 | 0.060 | 0.046 | 0.046 | 0.063 | 0.063 | 0.052 | 0.050 |
| 1000 | 0.051 | 0.052 | 0.046 | 0.049 | 0.054 | 0.053 | 0.046 | 0.045 | 0.058 | 0.056 | 0.053 | 0.051 |
| | | | | | MAF=0.2 | | | | | | | |
| 20 | 0.072 | 0.096 | 0.041 | 0.040 | 0.071 | 0.089 | 0.037 | 0.037 | 0.083 | 0.106 | 0.050 | 0.049 |
| 50 | 0.066 | 0.064 | 0.043 | 0.040 | 0.074 | 0.070 | 0.043 | 0.041 | 0.081 | 0.077 | 0.046 | 0.045 |
| 100 | 0.059 | 0.057 | 0.048 | 0.049 | 0.063 | 0.062 | 0.044 | 0.045 | 0.071 | 0.072 | 0.047 | 0.049 |
| 500 | 0.053 | 0.054 | 0.050 | 0.052 | 0.053 | 0.050 | 0.046 | 0.045 | 0.056 | 0.057 | 0.052 | 0.053 |
| 1000 | 0.052 | 0.051 | 0.049 | 0.049 | 0.051 | 0.050 | 0.050 | 0.047 | 0.056 | 0.051 | 0.052 | 0.050 |

## Table 3

**Type 1 error evaluation under simulation model 2 with 20% group uncertainty.** *Superscript $^{BG}$ denotes $TW_{LAD}$ and $gS_{LAD}$ being applied to the "best-guess" genotype data. The true genotype data were masked using a Dirichlet distribution for the genotype probabilities with scale parameters $a$ for the correct genotype and $(1-a)/2$ for the other two. On average $a = 0.8$ corresponds to 20% group uncertainty. See legend of Table 2 (main text) for additional simulation details.*

| $n/2$ | Gaussian | | | | Student's $t_4$ | | | | $\chi^2_4$ | | | |
|---|---|---|---|---|---|---|---|---|---|---|---|---|
| | $TW_{LAD}^{BG}$ | $TW_{LAD}$ | $gS_{LAD}^{BG}$ | $gS_{LAD}$ | $TW_{LAD}^{BG}$ | $TW_{LAD}$ | $gS_{LAD}^{BG}$ | $gS_{LAD}$ | $TW_{LAD}^{BG}$ | $TW_{LAD}$ | $gS_{LAD}^{BG}$ | $gS_{LAD}$ |
| | | | | | MAF=0.1 | | | | | | | |
| 20 | 0.082 | 0.100 | 0.039 | 0.036 | 0.082 | 0.102 | 0.039 | 0.040 | 0.096 | 0.117 | 0.048 | 0.048 |
| 50 | 0.080 | 0.074 | 0.045 | 0.046 | 0.093 | 0.081 | 0.046 | 0.049 | 0.105 | 0.093 | 0.047 | 0.049 |
| 100 | 0.064 | 0.062 | 0.048 | 0.050 | 0.084 | 0.075 | 0.049 | 0.050 | 0.090 | 0.082 | 0.050 | 0.051 |
| 500 | 0.049 | 0.052 | 0.046 | 0.047 | 0.062 | 0.060 | 0.051 | 0.052 | 0.060 | 0.060 | 0.049 | 0.050 |
| 1000 | 0.053 | 0.053 | 0.051 | 0.052 | 0.057 | 0.054 | 0.051 | 0.049 | 0.054 | 0.055 | 0.050 | 0.051 |
| | | | | | MAF=0.2 | | | | | | | |
| 20 | 0.063 | 0.073 | 0.038 | 0.040 | 0.069 | 0.075 | 0.041 | 0.042 | 0.082 | 0.083 | 0.049 | 0.048 |
| 50 | 0.062 | 0.057 | 0.043 | 0.043 | 0.065 | 0.057 | 0.041 | 0.042 | 0.071 | 0.068 | 0.046 | 0.048 |
| 100 | 0.053 | 0.053 | 0.046 | 0.047 | 0.062 | 0.060 | 0.046 | 0.047 | 0.065 | 0.062 | 0.046 | 0.046 |
| 500 | 0.049 | 0.046 | 0.044 | 0.044 | 0.055 | 0.056 | 0.051 | 0.052 | 0.054 | 0.055 | 0.051 | 0.052 |
| 1000 | 0.050 | 0.050 | 0.050 | 0.050 | 0.050 | 0.050 | 0.048 | 0.050 | 0.052 | 0.054 | 0.050 | 0.050 |

**Table 4**

*Type 1 error and power evaluation under simulation model 1 for singleton/pair combinations.*
Comparing $TW_{LAD}$ and $gS_{LAD}$ across various singleton/pair combinations. The design setup follows simulation model 1 with equal number of singletons and/or pairs in the MZ and DZ groups. See Section 3.1 for simulation model 1 details.

| $\sigma_1^2$ | $\sigma_2^2$ | $TW_{LAD}$ | $gS_{LAD}$ |
|---|---|---|---|
| | | 1000/0 Singletons/Pairs; n=1000 | |
| 1.0 | 1.0 | 0.053 | 0.053 |
| 1.1 | 1.0 | 0.169 | 0.168 |
| 1.2 | 1.0 | 0.476 | 0.475 |
| | | 900/100 Singletons/Pairs; n=1100 | |
| 1.0 | 1.0 | 0.051 | 0.051 |
| 1.1 | 1.0 | 0.169 | 0.170 |
| 1.2 | 1.0 | 0.481 | 0.484 |
| | | 0/1000 Singletons/Pairs; n=2000 | |
| 1.0 | 1.0 | 0.053 | 0.053 |
| 1.1 | 1.0 | 0.223 | 0.224 |
| 1.2 | 1.0 | 0.628 | 0.628 |

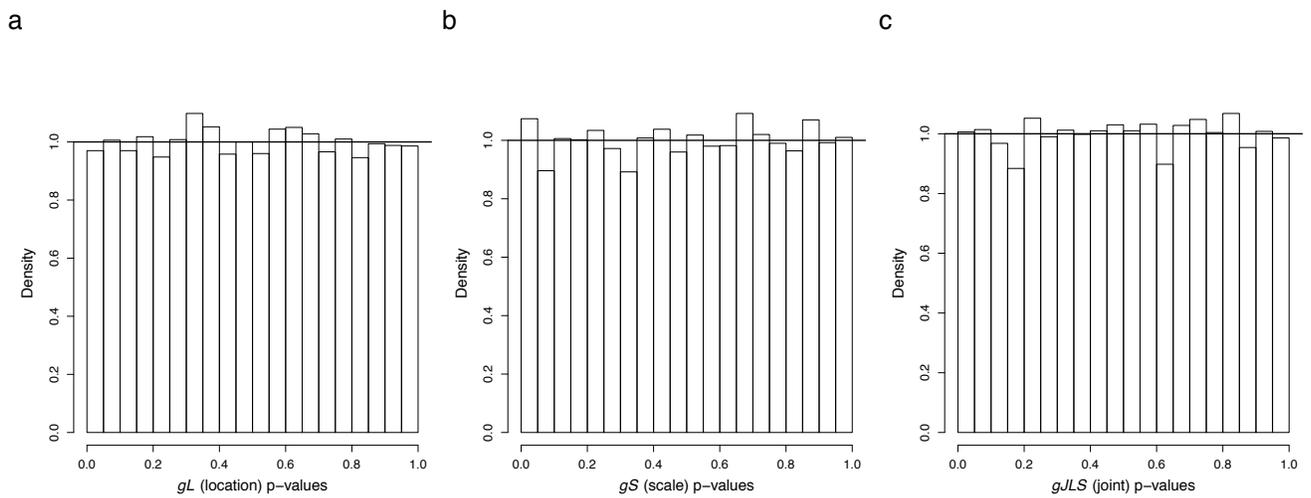

**Figure 1. Empirical Type 1 error of generalized tests (location, scale and joint) for CF lung disease dataset.** 10,000 permutation replicates of the lung function measures were generated from the full sample of Cystic Fibrosis patients including siblings ($n$=1507). For each replicate, the rs17563161 SNP genotype ($SLC9A3$) was analyzed directly using the generalized location ([a] $gL$), scale ([b] $gS$), and joint ([c] $gJLS$) tests. The location and scale tests were done under additive [df=1] and genotypic [df=2] genetic models, respectively. Resulting $p$-values are plotted as histograms and tested for deviation from a uniform distribution using the Kolmogorov-Smirnov test (p=0.71, 0.43 and 0.55 for (a), (b) and (c), respectively). See Section 4.2 for additional details.

}

\end{document}